\documentclass[aps,prl,superscriptaddress,longbibliography,showpacs,twocolumn]{revtex4-2}

\usepackage{pdfpages}

\makeatletter
\AtBeginDocument{\let\LS@rot\@undefined}
\makeatother

\usepackage{soul}
\setstcolor{red}
\usepackage{comment}

\newcommand{\tr}{\mathrm{tr}}

\usepackage{lipsum}
\usepackage{caption}
\DeclareCaptionJustification{justified}{\leftskip=0pt \rightskip=0pt \parfillskip=0pt plus 1fil}

\usepackage{graphicx}% Include figure files
\usepackage{dcolumn}% Align table columns on decimal point
\usepackage{bm}% bold math
\usepackage{caption}
\usepackage{amsmath,amssymb,amsthm}
\usepackage{braket}
\usepackage{empheq}
\usepackage{subfig}
\usepackage{tikz}
\usepackage{url}
\usepackage{hyperref}
\usepackage{quantikz}
\usepackage{xcolor}
\usepackage{float}

\usepackage{algorithm}
\usepackage[noend]{algpseudocode}% http://ctan.org/pkg/algorithmicx

\hypersetup{
    breaklinks=true,
    colorlinks=true,       % false: boxed links; true: colored links
    linkcolor=blue,          % color of internal links
    citecolor=red,        % color of links to bibliography
    filecolor=magenta,      % color of file links
    urlcolor=blue,           % color of external links
    runcolor=cyan
}

\newtheorem{theorem}{Theorem}
\newtheorem*{theorem*}{Theorem}
\newtheorem{corollary}{Corollary}%[theorem]

\usepackage{etoolbox}
\makeatletter
\patchcmd{\@citex}{\unskip, et al.}{\unskip et al.}{}{}
\makeatother

\begin{document}

\title{Faster Randomized Dynamical Decoupling}
\author{Changhao Yi}
\affiliation{Center for Quantum Information and Control, University of New Mexico, NM 87131, USA}
\affiliation{Department of Physics and Astronomy, University of New Mexico, NM 87131, USA}
\affiliation{State Key Laboratory of Surface Physics and Department of Physics, Fudan University, Shanghai 200433, China}
\affiliation{Institute for Nanoelectronic Devices and Quantum Computing, Fudan University, Shanghai 200433, China}
\author{Leeseok Kim}
\affiliation{Center for Quantum Information and Control, University of New Mexico, NM 87131, USA}
\affiliation{Department of Electrical and Computer Engineering, University of New Mexico, NM 87131, USA}
\author{Milad Marvian}
\affiliation{Center for Quantum Information and Control, University of New Mexico, NM 87131, USA}
\affiliation{Department of Physics and Astronomy, University of New Mexico, NM 87131, USA}
\affiliation{Department of Electrical and Computer Engineering, University of New Mexico, NM 87131, USA}

\begin{abstract}
We present a randomized dynamical decoupling (DD) protocol that can substantially improve the performance of any given deterministic DD, by using no more than two additional pulses. 
Our construction is implemented by probabilistically applying sequences of pulses, which, when combined, effectively eliminate the error terms that scale linearly with the system-environment coupling strength. 
As a result, we show that a randomized protocol using a few pulses can outperform deterministic DD protocols that require considerably more pulses.
Furthermore, we prove that the randomized protocol provides an improvement compared to deterministic DD sequences that aim to reduce the error in the system's Hilbert space, such as Uhrig DD, which had been previously regarded to be optimal. To rigorously evaluate the performance, we introduce new analytical methods suitable for analyzing higher-order DD protocols that might be of independent interest. We also present numerical simulations confirming the significant advantage of using randomized protocols compared to widely used deterministic protocols.

\end{abstract}

\maketitle

Overcoming decoherence is crucial for the development of reliable quantum technologies, as it directly affects the system’s ability to preserve quantum information over time. A powerful technique for addressing this issue is dynamical decoupling (DD)  \cite{viola1999dynamical,viola1998dynamical,vitali1999using,zanardi1999symmetrizing}, which applies fast control pulse sequences to individual qubits to average out system-environment interactions. Due to its simplicity in concept, design, and implementation, DD has been applied across a wide range of experimental platforms, including superconducting devices \cite{bylander2011noise,pokharel2018demonstration,jurcevic2021demonstration,tripathi2022suppression,ezzell2023dynamical}, trapped ions \cite{biercuk2009experimental,biercuk2009optimized}, solid-state spins \cite{lange2010universal,farfurnik2015optimizing, wang2012comparison,choi2020robust,zhou2024robust}, allowing for state-of-the-art quantum computing demonstrations on various platforms \cite{acharya2024quantum,acharya2023suppressing,kim2023scalable,kim2023evidence, bluvstein2024suppressing,xu2024constant,evered2023high}.

Over the years, various DD methods have been developed. These range from basic pulse sequences like Carr-Purcell-Meiboom-Gill \cite{meiboom1958modified} and XY4 \cite{maudsley1986modified}, which eliminates single-qubit decoherence to the low order of the time between pulses. However, this pulse interval cannot be reduced arbitrarily small due to experimental constraints, which limits the effectiveness of basic DD schemes. To suppress the system-environment interaction more effectively, advanced protocols such as concatenated DD \cite{khodjasteh2005fault}, Uhrig DD \cite{uhrig2007keeping}, and quadratic DD \cite{west2010near} have been developed. Another important family of DD protocols, is pulse-randomized DD \cite{viola2005random,santos2006enhanced}. Unlike deterministic DD methods mentioned above that use a fixed sequence of pulses, the pulse-randomized DD selects each pulse randomly at each time step from the set of pulses used in deterministic approaches. This randomized protocol is shown to be more effective at suppressing errors over long evolution times and more robust against time-dependent system uncertainties compared to deterministic counterparts \cite{santos2006enhanced,santos2008advantages}. Importantly, however, they do not improve the error scaling. {Recently, it has been shown that random timings of pulses  can  enable the measurement of arbitrary linear functionals of the noise spectrum \cite{huang2024random}.}

In this Letter, inspired by recent advancements in Hamiltonian simulation \cite{childs2019faster,campbell2017shorter,hastings2017turning,campbell2019random}, we propose a sequence-randomized DD method that  improves the error scaling of existing deterministic DD methods while using at most two additional pulses. %The proposed sequence-randomized method is fundamentally different from previously discussed pulse-randomized DD approaches, as it selects a \textit{pulse sequence} randomly from a set of pulse sequences. 
We show that for any given deterministic DD method, there exists a corresponding set of \textit{pulse sequences} such that choosing sequences uniformly at random from this set improves the error scaling. %\so{with respect to both the pulse interval $\tau$ and the system-environment coupling strength $J$.} 
Specifically, we prove that the sequence-randomized DD entirely removes the error term that scales linearly with system-environment coupling strength $J$, unlike any deterministic DD protocols. Consequently, we show that a first-order sequence-randomized protocol can outperform higher-order deterministic counterparts in the weak-coupling regime, in some cases leading to an exponential reduction in the number of pulses. 

The advantage of the sequence-randomized protocol is even more significant when focusing on the error restricted to the system's Hilbert space. We prove that, for any DD protocol targeting errors within the system Hilbert space, the randomized protocol can provide a quadratic improvement. For instance, it can surpass the error scaling of Uhrig DD (UDD) \cite{uhrig2007keeping}, which is a widely used DD method that suppresses dephasing noise. UDD had been considered to be \textit{optimal}, as it reduces one additional order of error in $T$ with one extra pulse where $T$ is the total time of the sequence. We prove that our sequence-randomized version of UDD reduces \textit{two} additional orders of $T$ with one extra pulse, significantly surpassing the performance of deterministic UDD. Specifically, the deterministic UDD reduces the error to $\mathcal{O}(T^{K+1})$ using $K$ pulses, whereas the randomized UDD reduces the error to $\mathcal{O}(T^{2K+2})$ using at most $K+2$ pulses. Therefore, for any $K > 3$, the randomized UDD outperforms the deterministic counterpart with the same number of pulses. We show that a similar advantage can be achieved in other widely used DD protocols such as quadratic DD \cite{west2010near}. 

We present numerical simulations that confirm the substantial improvement of the proposed methods, with gains of many orders of magnitude in some cases, across a wide range of experimentally relevant parameters, and requiring significantly fewer pulses compared to currently common DD sequences.
To provide rigorous performance bounds for high-order DD protocols, we introduce the generalized Relative Action Integral (RIA) method \cite{bookatz2015error,marvian2017error,burgarth2022one}, which bounds all the high-order error terms and does not rely on convergence assumptions. This is in contrast to approaches based on Magnus expansion and average Hamiltonian theory \cite{blanes2009magnus} which rely on convergence assumptions that may not align with many practical experimental conditions \cite{ oon2024average, wang2012comparison}. Therefore we expect the new RIA method to be of independent interest in analyzing high-order DDs.

\textit{Illustrative Example.}--- Consider a quantum system $S$ coupled to an arbitrary bath $B$, defined by Hilbert spaces $\mathcal{H}_S$ and $\mathcal{H}_B$, respectively. The total Hamiltonian can be written as $H = H_S \otimes I_B + I_S \otimes H_B + H_{SB}$, where $H_S$ and $H_B$ correspond to the system and bath Hamiltonians, respectively, while $H_{SB}$ describes the interaction between the system and the bath. In order to protect the evolution of $S$ from the effect of $H_{SB}$, DD \cite{viola1999dynamical,viola1998dynamical,vitali1999using,zanardi1999symmetrizing} is achieved by introducing a time-dependent control Hamiltonian $H_C(t)$ that acts only on the system Hilbert space $\mathcal{H}_S$, which generates a sequence of ideal pulses at desired time intervals.

As a concrete example, consider an $n$-qubit 1D Heisenberg spin chain $H_S = \sum_{j=1}^n (X_j X_{j+1} + Y_j Y_{j+1} + Z_j Z_{j+1})$ interacting with a qubit-bath that introduces local dephasing noise to the system. The system-bath interaction is described by $H_{SB} = J \sum_{j=1}^n Z_j \otimes B_{j}$ where $B_{j}$ is a general bath operator, and the bath Hamiltonian $H_B$ is chosen as an arbitrary operator. If we choose the control Hamiltonian as $H_C(t) = \frac{\pi}{2}H_X\left(\delta(t ) + \delta(t - \tau)\right),$ with $H_X = \sum_{j=1}^n X_j$, the effect of $H_C(t)$ is applying two $X^{\otimes n}$ pulses on the system at times $t=0$ and $t=\tau$ (which is the parallel application of  spin or Hahn echo \cite{hahn1950spin} on each qubit). The total evolution at time $t=2\tau$ becomes:
\begin{align}
    D = e^{-iH\tau}X^{\otimes n} e^{-iH\tau} X^{\otimes n}.
\end{align} Such a pulse sequence preserves $H_S$ while reversing the sign of $H_{SB}$ for half of the time, effectively averaging out its effect. In fact, for any given initial state $\rho$, the DD-protected state $D\rho D^\dagger$ approximates the ideal state (evolved with $H_{SB} = 0$) by
\begin{align}\label{example_errorbound}
    \Vert D\rho D^\dagger - e^{-i 2H_0 \tau} \rho e^{i2 H_0 \tau}\Vert_1 = \mathcal{O}(J\tau^2),
\end{align} where $H_0 = H_S + H_B$. (See
the Supplemental Material (SM) for the explicit derivations of this example.) 

Instead, we can also consider a similar deterministic protocol where the position of the first pulse $X^{\otimes n}$ is shifted to the end of the sequence $t=2\tau$:
\begin{align}
    D^{\text{rev}} = X^{\otimes n}e^{-iH\tau}X^{\otimes n} e^{-iH\tau}.
\end{align} 
Clearly $D^{\text{rev}}$ achieves the same error scaling as in Eq.\eqref{example_errorbound}.

Our proposed randomized scheme (uniformly) randomly chooses between the two previous deterministic sequences: with equal probability we either apply the two pulses at $t=0,\tau$, or apply them at $t=\tau, 2\tau$. The state after implementing this mixed unitary channel  is:
\begin{align}
    \mathcal{R}(\rho) = \frac{1}{2} D \rho D^\dagger + \frac{1}{2} D^{\text{rev}} \rho {D^{\text{rev}}}^\dagger.
\end{align} 
Remarkably, such sequence-randomized DD yields a significantly reduced error scaling of
\begin{align}\label{example_errorbound}
    \Vert \mathcal{R}(\rho) - e^{-i2 H_0\tau} \rho e^{i2H_0\tau}\Vert_1 = \mathcal{O}(J^2\tau^3),
\end{align} which demonstrates the scaling advantage in \textit{both} $J$ and $\tau$ using the same number of pulses and only by introducing simple classical randomness. Since the protocol removes the first-order terms in $J$ completely, this simple analysis shows that this probabilistic protocol can beat any higher-order DD sequence that scales as $\mathcal{O}(J\tau^K)$, for any $K$, in the weak coupling regime. Here the particular choice of the random DD protocol makes it possible to map the evolution to that of the randomized first-order product formula presented in Ref.\cite{childs2019faster}. Note that the channel  $\mathcal{R}$ can be implemented by randomly choosing and implementing one DD sequence in each compilation and therefore incurs no additional sampling costs when estimating the expectation values of observables. The superior performance of the randomized protocols also suggests the potential of novel deterministic protocols where several carefully designed deterministic DD sequences are implemented and their results are averaged out to reduce the effect of noise in estimation (see SM for de-randomized examples). In what follows we generalize this method and provide rigorous error bounds.

\textit{General Construction.}--- We consider the decoupling group $G$ where $g_\ell \in G$ acts on the system Hilbert space $\mathcal{H}_S$. The  application of pulses $g_0^\dagger, \dots, g_{\ell+1}^\dagger g_\ell,\dots , g_{L-1}$ at $\tau$ intervals generates 
\begin{equation}\label{deterministic_protocol}
    D = \prod_{\ell = 0}^{L-1}g_\ell \exp[-i (H_0 +  H_{SB} )\tau]g_\ell^\dagger,\quad g_\ell \in G,
\end{equation} 
where $H_0 = H_S + H_B$. (For simplicity we consider equal intervals in this section, but will discuss sequences with  different intervals later.) The decoupling group $G$ is chosen to ensure that they commute with the system Hamiltonian but eliminate the effect of interaction with environment \cite{viola1998dynamical,viola1999dynamical,vitali1999using,zanardi1999symmetrizing}
\begin{equation}\label{equ:1storder}
    \forall \ell, \, [g_\ell,H_0] = 0,\quad \sum_{\ell=0}^{L-1}g_\ell H_{SB} g_\ell^\dagger= 0.
\end{equation} 
A DD sequence achieves a $K$-th order decoupling if  $\|D - U_0\| = \mathcal{O}(J \tau^{K+1}) + \mathcal{O}(J^2)$ \cite{ng2011combining,khodjasteh2007performance}, where $J=\|H_{SB}\|$ is the system-environment interaction strength. Such high-order DD sequences can be constructed at the expense of increasing the number of pulses and the total time of the sequence $T=L\tau$. Significant effort has been put into designing advanced DD sequences aiming to improve the error bound \cite{khodjasteh2005fault,uhrig2007keeping,west2010near,wang2011protection}, for example, concatenated DD (CDD) which uses exponential in $K$ many pulses to achieve this error scaling \cite{khodjasteh2005fault,khodjasteh2007performance,ng2011combining}.

Given any such a deterministic DD sequence, to implement the sequence-randomized DD protocol we (uniformly) randomly choose one element of the decoupling group  $g\in G$, and only modify the first and last pulse, i.e., implementing $g_0^\dagger g^\dagger$ instead of  $g_0^\dagger$ and implementing $g g_{L-1}$ instead of $g_{L-1}$ respectively, while keeping the rest of the pulses unchanged. The corresponding quantum channel is described by 
\begin{equation}\label{randomized_protocol}
    \mathcal{R}(\rho) = \frac{1}{|G|}\sum_{g\in G} (g D g^\dagger)\rho (gD^\dagger g^\dagger).
\end{equation} 
Let $\mathcal{U}_0(\rho) = U_0 \rho U_0^\dagger$ denote the quantum channel representing the ideal evolution, and $U_0=e^{-i H_0 T}$. We derive the following error bound for the sequence-randomized DD protocol for a given deterministic DD protocol:

\begin{theorem}\label{Theorem1}
    Consider any $K$-th order deterministic dynamical decoupling protocol $D$ which satisfies 
    \begin{align}
        \|D - U_0\| = \mathcal{O}(J \tau^{K+1}) + \mathcal{O}( J^2).
    \end{align} Then, the sequence-randomized protocol given by Eq.\eqref{randomized_protocol} yields
    \begin{align}
        \|\mathcal{R} - \mathcal{U}_0\|_\diamond = \|D - U_0\|^2 + \mathcal{O}(J^2)=\mathcal{O}(J^2).
    \end{align}
\end{theorem} 
The proof with precise (but lengthy) upper bounds are available in the SM. The statement of  Theorem~\ref{Theorem1} emphasizes two significant results. First, our sequence-randomized DD protocol completely eliminates error terms that scale linearly with respect to the system-environment interaction strength $J$. Therefore, in the weak coupling regime, the randomized protocol always achieves better error scaling compared to its deterministic counterpart (for any order $K$). Even more strikingly,  in the weak coupling regime, a randomized \textit{first-order} DD protocol can potentially outperform any $K$-th order deterministic DD! 

Second, given the significant improvement in the performance of the randomized protocol and elimination of first-order $J$ errors, rigorous analysis of higher-order error terms becomes more important to capture the remaining errors. To do so, we introduce a generalization of the Relative Integral Action \cite{bookatz2015error,marvian2017error,burgarth2022one} and find new rigorous bounds on the performance of higher-order DD sequences. 
%\blue{do we need to say that the breakdown of the convergence condition here again?} MM: No

As a physically relevant example, consider a local bath that introduces general 1-local noise to each system qubit \cite{ng2011combining}. 
The decoupling group that can suppress such 1-local noise is $G = \{I^{\otimes n}, X^{\otimes n}, Y^{\otimes n}, Z^{\otimes n}\}$. The corresponding sequence is known as universal decoupling or XY4 \cite{maudsley1986modified}, applied to every qubit in parallel. A simple recursive application of such sequences $K-1$ times, known as concatenated DD \cite{khodjasteh2005fault} or CDD$_K$, can achieve an error $\mathcal{O}(J\tau^{K+1})$ by using $4^{K}$ number of pulses with equal pulse interval $ \tau$, i.e. $T = 4^{K}\tau$. The performance of CDD has been rigorously analyzed using the Magnus expansion \cite{ng2011combining}. An application of the generalized RIA method and  Theorem~\ref{Theorem1} for randomized CDD provides the following performance bounds.

\textit{Example}---
Consider the Hamiltonian $H = H_0 + H_{SB}$ where  $H_{SB} = \sum_{j=1}^n X_j\otimes B_{X,j} + Y_j\otimes B_{Y,j} + Z_j\otimes B_{Z,j}$, and $J = \sum_{j=1}^n \|B_{X,j}\| + \|B_{Y,j}\| + \|B_{Z,j}\|$, $\beta=\|H_{0}\|$. The deterministic CDD$_K$ protocol yields the error bound: 
\begin{align}\label{CDD_deterministic}
    \|D - U_0\| \le T(2\beta)^{K} &J2^{K(K-1)+1}\tau^K \\
    &+ J^2T\left[2 + T (2\beta  + J)\right]\tau. \nonumber 
\end{align}
In contrast, applying Theorem~\ref{Theorem1}, the first-order sequence-randomized DD protocol  has error bound
\begin{equation}\label{CDD_randomized}
    \|\mathcal{R} - \mathcal{U}_0\|_{\diamond} \leq \Vert D - U_0 \Vert^2 + J^2T[2+T(2\beta+J)]\tau,
\end{equation}
which exhibits the advantage, as linear terms in $J$ are removed. 
Note that our bound on the diamond distance also bounds the trace distance for any input state. The diamond distance and trace distance have precise operational interpretations in distinguishing quantum channels and quantum states, respectively \cite{watrous2018the}. In contrast, other measures, such as infidelity,  can be (quadratically) looser than the trace distance in distinguishing mixed states (see SM).

\textit{Quadratic Improvement in Subsystem Error}---
For DD protocols aimed at reducing errors only within the system Hilbert space, such as Uhrig DD \cite{uhrig2007keeping}, quadratic DD \cite{west2010near}, and their variants \cite{mukhtar2010protecting,wang2011protection}, randomization offers  quadratic enhancement in performance. Let $H =  I_S\otimes H_B + H_{SB}$. 
For any initial state of the form $\rho = \rho_S \otimes \rho_B \in \mathcal{H}_S \otimes \mathcal{H}_B$, the reduced state in the system Hilbert space after applying the deterministic DD sequence can be expressed as $\rho_S(T) = \tr_B(D \rho D^\dagger)$, where $D$ represents the deterministic DD protocol described in Eq.\eqref{deterministic_protocol}. 
Denote the reduced state after the application of the randomized DD protocol described by $\mathcal{R}(\rho)$ in Eq.\eqref{randomized_protocol}  by $\rho_S^{\text{ran}}(T) = \tr_B(\mathcal{R}(\rho))$. The subsystem errors of interest are $\frac{1}{2}\Vert \rho_S(T) - \rho_S(0) \Vert_1$ and $\frac{1}{2}\Vert \rho_S^{\text{ran}}(T) - \rho_S(0) \Vert_1$.

Consider the expansion of a deterministic DD protocol $D$ into (multi-qubit) Pauli operators $\mathcal{P} = \{P\}$ in the system Hilbert space: $D = \sum_{P \in \mathcal{P}} P\otimes E_P$. Assuming $\sum_{\mathcal{P}\setminus\{I\}} \Vert E_P \Vert \leq 1$, it is straightforward to show that \cite{uhrig2010rigorous,kuo2011quadratic} (also see SM)
    \begin{align} \label{eq:subsystem_deter}
        \frac{1}{2}\Vert \rho_S(T) - \rho_S(0) \Vert_1 \leq 2 \sum_{\mathcal{P}\setminus\{I\}} \Vert E_P\Vert.
    \end{align} 
 The terms $\{E_P\}$ (for $\mathcal{P}\setminus\{I\}$) can be interpreted as the strength of the environment's non-trivial effect to the system's evolution. In fact, if $E_P = 0$ for all $P\in\mathcal{P}\setminus\{I\}$, then the DD protocol yields $D = I \otimes E_I$, so that tracing out the environment leaves the state unchanged. Hence, many existing DD protocols \cite{uhrig2007keeping,yang2008universality,west2010near,mukhtar2010protecting,uhrig2010rigorous,kuo2011quadratic,wang2011protection} aim to suppress $\sum_{\mathcal{P}\setminus\{I\}} \Vert E_P \Vert$. 
 
\begin{theorem}
\label{Theorem2}
Consider the expansion of a given deterministic DD protocol $D=\sum_{P \in \mathcal{P}} P\otimes E_P$, where $\mathcal{P}$ is the set of required (multi-qubit) Pauli operators on the system Hilbert space,  satisfying $\sum_{\mathcal{P}\setminus\{I\}} \Vert E_P \Vert \leq 1$. Choose a decoupling group $G$ that satisfies $
   \forall P \in \mathcal{P}: \,  \sum_{g\in G} g P g^\dagger = 0$ and define the corresponding sequence-randomized DD as $\mathcal{R}(\rho) = \frac{1}{|G|}\sum_{g\in G} (g D g^\dagger)\rho (gD^\dagger g^\dagger)$. Then we have
    \begin{align}
    \frac{1}{2}\Vert \rho_S^{\text{ran}}(T) - \rho_S(0) \Vert_1 \leq 2 \left(\sum_{\mathcal{P}\setminus\{I\}} \Vert E_P\Vert\right)^2, \quad  \forall \rho_S(0).
    \end{align}
\end{theorem}

\begin{figure}[t!]
    \centering
    \includegraphics[width=8.6cm]{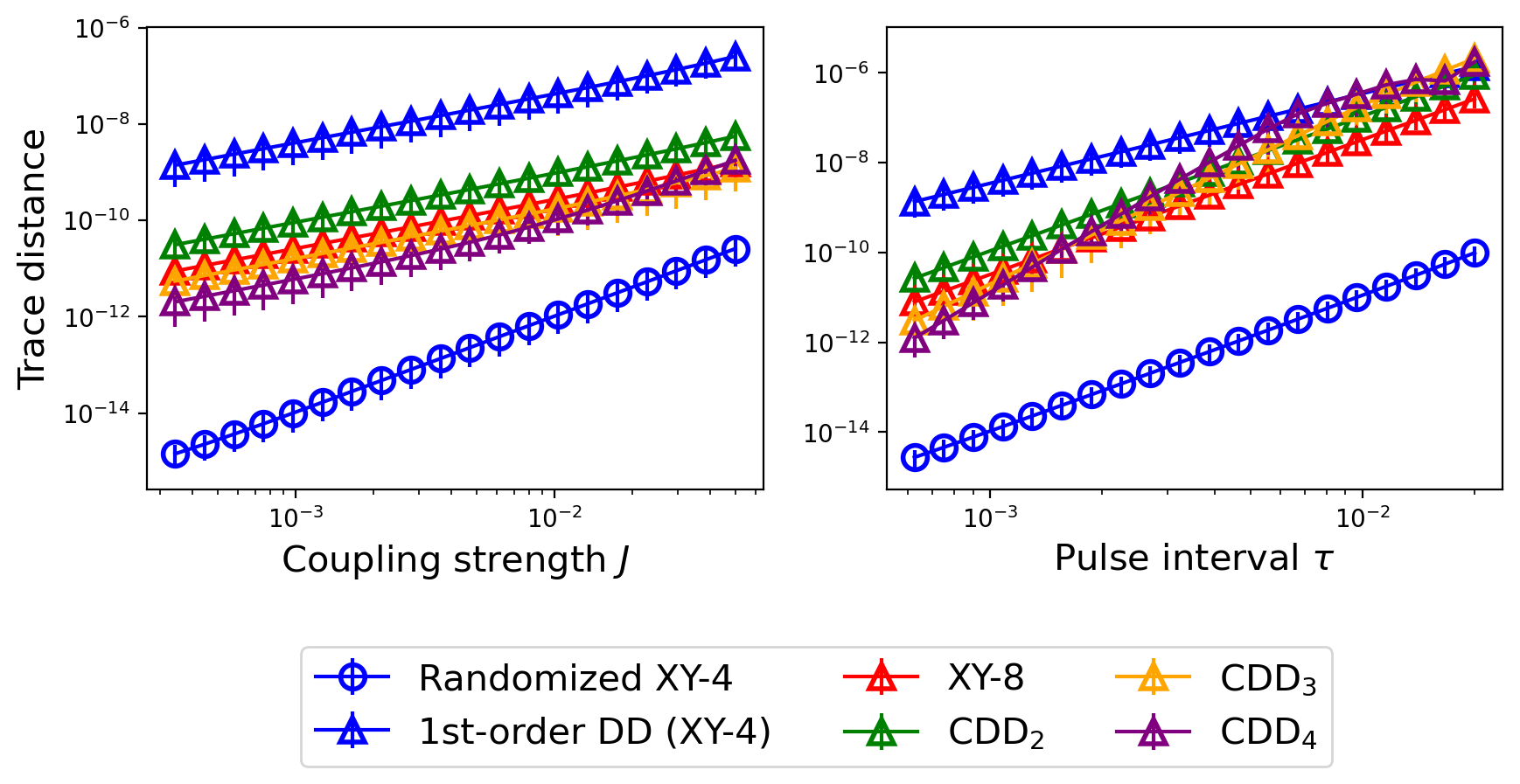}
    \caption{Performance of the sequence-randomized XY4,  represented by circles compared to deterministic protocols (CDD$_K$ for $K=1,2,3,4$, where CDD$_1=$ XY4,  and XY8), represented by triangles. Left panel: Fixed $\tau = 10^{-3}$ while varying $J$. Randomized XY4 protocol scales $\mathcal{O}(J^2)$, outperforming all deterministic protocols that scale as $\mathcal{O}(J)$. Right panel: Fixed $J = 10^{-3}$ while varying $\tau$. Note that CDD$_K$ requires $4^K$ many pulses but the randomized XY4 sequence uses no more than $6$ pulses.}
    \label{fig:fig1}
\end{figure} 

Therefore the randomized protocol provides a quadratic improvement compared to the bound for the deterministic protocol $D$ in Eq.~\eqref{eq:subsystem_deter}.
As an important example, consider Uhrig DD (UDD) \cite{uhrig2007keeping}, which is a widely used DD scheme to suppress dephasing. In this case, the system-bath Hamiltonian is $H = I_S \otimes B_I + Z \otimes B_Z$. It has been shown that by applying only $K$ pulses at specific times over the total time $T$, the UDD protocol can be expressed as $D = I_S \otimes E_I + Z \otimes E_Z$ with $\Vert E_Z \Vert = \mathcal{O}(T^{K+1})$ \cite{yang2008universality}. Therefore Eq.~\eqref{eq:subsystem_deter} guarantees that $\frac{1}{2}\Vert \rho_S(T) - \rho_S(0) \Vert_1 = \mathcal{O}(T^{K+1})$ \cite{uhrig2010rigorous}. Since adding an extra pulse reduces the error by one order of magnitude, UDD has been regarded as optimal.  To construct the sequence-randomized protocol, we note that $\mathcal{P}=\{I,Z\}$, and therefore we choose the decoupling group to be $G = \{I, X\}$. Hence the randomized UDD is described by:
\begin{align}
    \rho_S^{\text{ran}}(T) = \tr_B\left(\frac{1}{2}D \rho D^\dagger + \frac{1}{2}XDX \rho XD^\dagger X\right).
\end{align} 
Applying Theorem \ref{Theorem2} shows that in our proposed scheme adding each pulse can remove two orders of $T$.
\begin{corollary}\label{corollary2}

The randomized UDD  constructed from a $K$-th order UDD achieves
\begin{align}
    \frac{1}{2}\Vert \rho_S^{\mathrm{ran}}(T) - \rho_S(0) \Vert_1 = \mathcal{O}(T^{2K+2}),  \quad  \forall \rho_S(0).
\end{align}
\end{corollary} 
Note that the randomized UDD uses at most $K+2$ pulses to achieve $\mathcal{O}(T^{2K+2})$ error scaling, due to the two additional $X$ pulses in the protocol. Regardless, using the same number of pulses, the randomized UDD significantly outperforms the deterministic UDD for any $K > 3$. A randomization protocol can similarly improve Quadratic DD (QDD) \cite{west2010near}, which suppresses general 1-qubit noise (see SM).

\begin{figure}[t!]
    \centering
    \includegraphics[width=8.6cm]{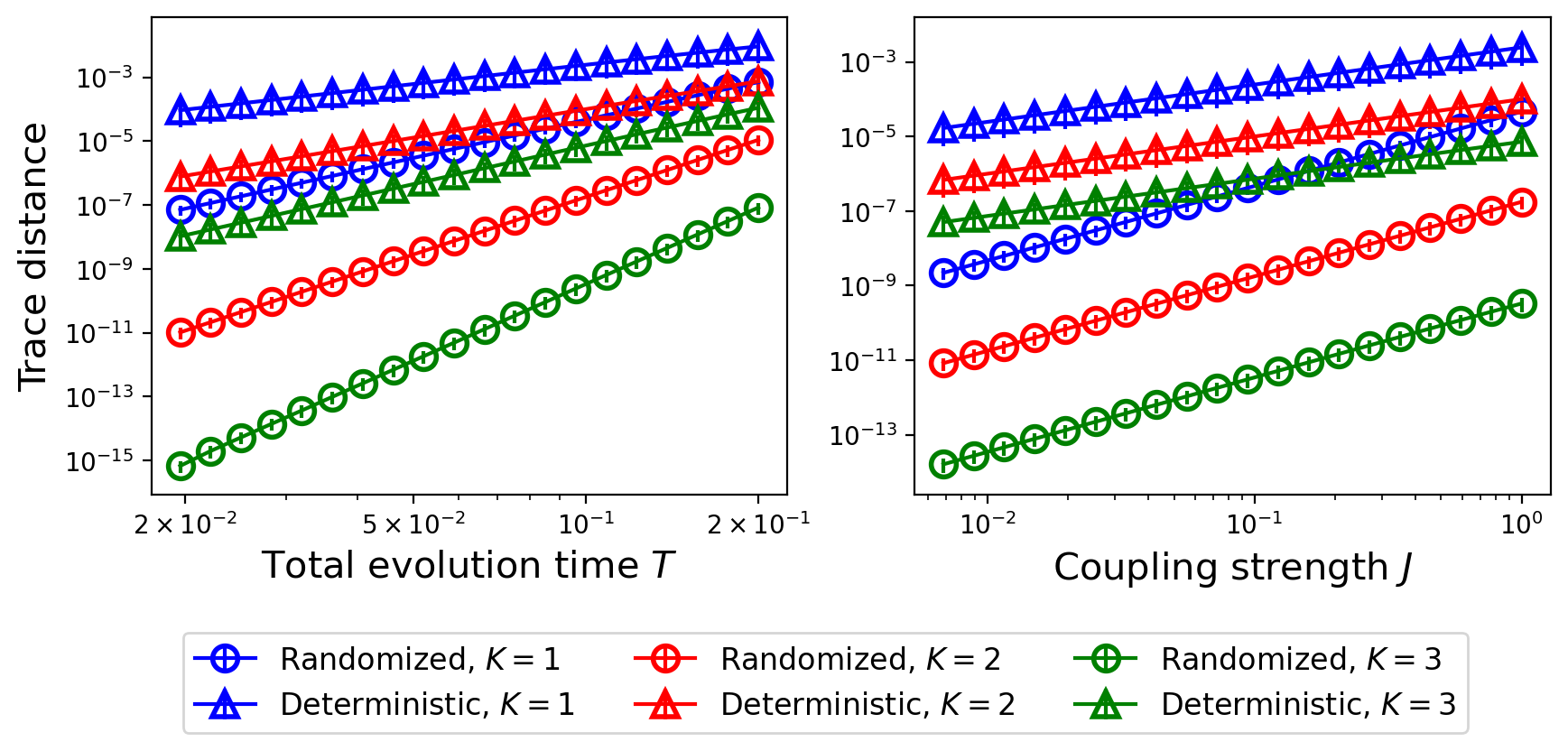}
    \caption{Performance of deterministic UDD for $K=1,2,3$ order and their corresponding sequence-randomized versions. Left panel: Fixed $J = 1$ while varying total evolution time $T$. Randomized DD achieves quadratic improvement, e.g., the slope of the first-order randomized UDD matches that of the third-order deterministic UDD. Right panel: Fixed $T=0.1$ while varying $J$. Randomized UDD does not exhibit $\mathcal{O}(J)$ error scaling observed in the deterministic case.}
    \label{fig:UDD}
\end{figure}

\textit{Numerical simulations.}--- We now present numerical results demonstrating the effectiveness of our sequence-randomized DD method for preserving arbitrary initial quantum states $\rho = \rho_S \otimes \rho_B$ where $\rho_S$ and $\rho_B$ are randomly chosen pure states, and the results are averaged over $20$ random initial states. (See SM for additional numerical results.) 
We first consider the 4-qubit Heisenberg spin chain $H_S = \sum_{j=1}^3 (X_j X_{j+1} + Y_j Y_{j+1} + Z_j Z_{j+1})$, which is subject to general $1$-local noise induced by the $4$-qubit bath, i.e., $H_{SB} = J \sum_{i=1}^4 \sum_{\alpha} \sigma_{\alpha, i} \otimes B_{\alpha}, H_B = I \otimes \sum_{i=1}^4 B_{I'}$ ($\sigma_{\alpha, i} \in \{X_i, Y_i, Z_i, I_i\}$) where $B_{\alpha} = \sum_{j=1}^4 \sum_{\beta} \gamma_{\alpha, \beta, j} \sigma_{\beta, j}$ with coefficients $\gamma_{\alpha,\beta,j}$ randomly selected from $[0,1]$. As previously discussed, the decoupling group $G = \{I, X^{\otimes n}, Y^{\otimes n}, Z^{\otimes n}\}$ eliminates $H_{SB}$ to first order in $\tau$. The corresponding pulse sequence is known as XY4 \cite{meiboom1958modified} or universal decoupling (applied to every qubit in parallel), which we denote by $D$. Our randomized protocol $\mathcal{R}$, as defined in Eq.\eqref{randomized_protocol}, averages over these group elements.

Fig.\ref{fig:fig1} compares the sequence-randomized XY4 (universal decoupling) protocol to deterministic CDD$_K$ protocols for $K =1,2,3,4$, and XY8 sequence which is a symmetrized version of XY4. The left panel shows the performance as a function of the coupling strength $J$, with fixed pulse interval $\tau = 10^{-3}$. As shown in Theorem \ref{Theorem1}, the randomized XY4 sequence scales as $\mathcal{O}(J^2)$, outperforming its deterministic counterparts that scale as $\mathcal{O}(J)$ for all orders $K$, in the weak-coupling regime. The right panel shows that the randomized XY4 still outperforms all the considered deterministic protocols for various pulse intervals $\tau$. This not only results in a significantly reduced error but also leads to a substantial reduction in the number of pulses, as CDD$_K$ uses $4^K$ pulses, while the randomized XY4 requires at most only $6$ pulses.

In Fig.\ref{fig:UDD} we present numerical simulations demonstrating the quadratic improvement achieved by randomizing Uhrig DD (UDD). We consider a 1-qubit system interacting with a 2-qubit environment, described by $H_{SB} = J(Z \otimes B_Z)$ and $H_B = I \otimes B_I$, where $B_k = \sum_{\alpha,\beta} \gamma_{k,\alpha,\beta} \sigma_{\alpha} \otimes \sigma_{\beta}$ and $\gamma_{k,\alpha,\beta}$ are randomly chosen from $[0,1]$ \cite{west2010near}. These bath operators include all 1- and 2-body terms. In the left panel of, fixing $J = 1$ we vary the total time of  UDD sequence $T$ and plot the trace distance for the deterministic UDD for $K=1,2,3$ order and their sequence-randomized versions. While the deterministic UDD sequences exhibit $\mathcal{O}(T^2)$, $\mathcal{O}(T^3)$, and $\mathcal{O}(T^4)$ error scalings, respectively, the corresponding randomized versions show $\mathcal{O}(T^4)$, $\mathcal{O}(T^6)$, and $\mathcal{O}(T^8)$ error scalings, which confirms Corollary \ref{corollary2}. The randomized protocol also does not have an $\mathcal{O}(J)$ term, as expected. In fact, numerical results show that randomized UDD significantly improves performance over deterministic counterparts for all orders, with the third-order randomized UDD improving the deterministic case by nearly eight orders of magnitude.

\textit{Conclusion.}--- We have introduced a new randomized DD protocol that improves the error scaling of the existing deterministic DD protocols by completely eliminating error terms that scale linearly with the system-environment coupling strength. Both rigorous error analysis and numerical simulations are presented confirming that the randomized protocol can significantly improve many widely used DD protocols such as Uhrig DD \cite{uhrig2007keeping}, which previously had been considered to be optimal.

Given the simplicity and effectiveness of our randomized protocol, we hope these results will inspire experimental implementations and further improve the performance of current quantum devices. 
Our findings open several interesting research directions as well. First is the possibility of designing protocols that can suppress the $\mathcal{O}(J^2)$ error terms. Another promising direction involves exploring the potential robustness against control errors and system uncertainties, as observed in pulse-randomized protocols \cite{viola2005random,santos2006enhanced,santos2008advantages}, as well as applications for  noise spectroscopy \cite{huang2024random} and quantum sensing \cite{wang2019randomization}. Finally, applications of our generalized RIA method for deriving tighter error bounds in related fields, such as quantum simulation, may be of independent interest.

\begin{acknowledgments}
\textit{Acknowledgments.}---%
This work is supported by DOE's Express: 2023 Exploratory Research For Extreme-scale Science Program under Award Number DE-SC0024685. Additional support by NSF
CAREER award No. CCF-2237356  is acknowledged.
\end{acknowledgments}

\bibliography{PRL_ref}



\section*{Supplementary material}

# Supplemental Material: Faster Randomized Dynamical Decoupling

Changhao Yi,[1, 2, 3, 4] Leeseok Kim,[1, 5] and Milad Marvian[1, 2, 5]

[1]*Center for Quantum Information and Control, University of New Mexico, NM 87131, USA*

[2]*Department of Physics and Astronomy, University of New Mexico, NM 87131, USA*

[3]*State Key Laboratory of Surface Physics and Department of Physics, Fudan University, Shanghai 200433, China*

[4]*Institute for Nanoelectronic Devices and Quantum Computing, Fudan University, Shanghai 200433, China*

[5]*Department of Electrical and Computer Engineering, University of New Mexico, NM 87131, USA*

## CONTENTS

## I. ILLUSTRATIVE EXAMPLE: TAYLOR EXPANSION

In the main context, we introduced an illustrative example of deterministic DD protocol $D$ and its sequence-randomized DD protocol. Although in the following sections we will develop rigorous techniques for general applications of the proposed scheme, it is instructive to use Taylor expansion to explain the main idea.

We denote the deterministic DD protocol as $\mathcal{D}(\rho) = D\rho D^\dagger$, while the randomized version is given by $\mathcal{R}(\rho) = \frac{1}{2}D\rho D^\dagger + \frac{1}{2}D^{\text{rev}}\rho D^{\text{rev}\dagger}$. Now we directly use linear expansion to demonstrate the advantage of randomization.

Let $H_0 = H_S + H_B$, $H_+ = H_0 + H_{SB}$ and $H_- = H_0 - H_{SB}$ with $\beta = \|H_0\|$ and $J = \|H_{SB}\|$. Using the identity

$$e^{-iH\tau}Ae^{iH\tau} = A - i\tau[H, A] + \frac{(-i\tau)^2}{2}[H, [H, A]] + \mathcal{O}(\tau^3) \tag{S1}$$

twice, it is straightforward to calculate $\mathcal{D}(\rho) = D\rho D^\dagger = e^{-iH_+\tau}e^{-iH_-\tau}\rho e^{iH_-\tau}e^{iH_+\tau}$ by expanding in powers of $\tau$:

$$\mathcal{O}(1) : \rho, \tag{S2}$$

$$\mathcal{O}(\tau) : -i\tau[H_-,\rho] - i\tau[H_+,\rho], \tag{S3}$$

$$\mathcal{O}(\tau^2) : \frac{(-i\tau)^2}{2}[H_+,[H_+,\rho]] + (-i\tau)^2[H_+,[H_-,\rho]] + \frac{(-i\tau)^2}{2}[H_-,[H_-,\rho]], \tag{S4}$$

$$\mathcal{O}(\tau^3) : \frac{(-i\tau)^3}{6}[H_-,[H_-,[H_-,\rho]]] + \frac{(-i\tau)^3}{6}[H_+,[H_+,[H_+,\rho]]]$$
$$+ \frac{(-i\tau)^3}{2}[H_+,[H_+,[H_-,\rho]]] + \frac{(-i\tau)^3}{2}[H_+,[H_-,[H_-,\rho]]]. \tag{S5}$$

If we replace $H_+$ and $H_-$ with $H_0$, we obtain the expansion for the ideal evolution $e^{-i2H_0\tau}\rho e^{i2H_0\tau}$. Hence, the difference between the two becomes

$$\mathcal{D}(\rho) - e^{-i2H_0\tau}\rho e^{i2H_0\tau} = (-i\tau)^2\left([H_{SB},[H_0,\rho]] + [H_0,[-H_{SB},\rho]]\right) + \mathcal{O}(\tau^3), \tag{S6}$$

$$= \mathcal{O}(\beta J\tau^2). \tag{S7}$$

Clearly the expression for $D^{\mathrm{rev}}\rho D^{\mathrm{rev}\dagger} - e^{-i2H_0\tau}\rho e^{i2H_0\tau}$ is similar to (S6) but with replacing $H_{SB}$ with $-H_{SB}$.

The key observation is that the resulting error in $\tau^2$ is exactly the negative of the second order term in Eq. (S6) since $H_{SB}$ shows up exactly once in each commutator:

$$D^{\mathrm{rev}}\rho D^{\mathrm{rev}\dagger} - e^{-i2H_0\tau}\rho e^{i2H_0\tau} = -(-i\tau)^2\left([H_{SB},[H_0,\rho]] + [H_0,[-H_{SB},\rho]]\right) + \mathcal{O}(\tau^3). \tag{S8}$$

To calculate $\mathcal{R}(\rho) - e^{-i2H_0\tau}\rho e^{i2H_0\tau}$ we can simply average the Eqs. (S6) and (S8) to see that the $\mathcal{O}(\tau^2)$ terms cancel out, i.e.:

$$\mathcal{R}(\rho) - e^{-i2H_0\tau}\rho e^{i2H_0\tau} = \mathcal{O}(\tau^3) \tag{S9}$$

as claimed.

To see the improvement in the scaling of $J$, an explicit calculation of $\mathcal{O}(\tau^3)$ term using Eq. (S7) gives

$$\frac{(-i\tau)^3}{3}\left([H_{SB},[H_0,[H_0,\rho]]] + [H_0,[H_{SB},[H_0,\rho]]] + [H_{SB},[H_{SB},[H_0,\rho]]]\right)$$
$$+ \frac{2(-i\tau)^3}{3}\left([H_{SB},[H_0,[H_0,\rho]]] + [H_0,[H_0,[-H_{SB},\rho]]] + [H_{SB},[H_0,[-H_{SB},\rho]]]\right)$$
$$+ \frac{(-i\tau)^3}{3}\left([H_0,[-H_{SB},[H_0,\rho]]] + [H_0,[H_0,[-H_{SB},\rho]]] + [H_0,[-H_{SB},[-H_{SB},\rho]]]\right). \tag{S10}$$

Averaging over $+H_{SB}$ and $-H_{SB}$ eliminates all terms containing an odd number of $H_{SB}$. Thus, the sequence-randomized DD protocol leads to an error of

$$\mathcal{R}(\rho) - e^{-i2H_0\tau}\rho e^{i2H_0\tau} = \frac{(-i\tau)^3}{3}\left([H_{SB},[H_{SB},[H_0,\rho]]] - 2[H_{SB},[H_0,[H_{SB},\rho]]] + [H_0,[H_{SB},[H_{SB},\rho]]]\right) + \mathcal{O}(\beta J^2\tau^4) \tag{S11}$$

$$= \mathcal{O}(\beta J^2\tau^3). \tag{S12}$$

Note that since every term that contains an odd number of $H_{SB}$ vanishes, the error contains only even orders of $J$.

## II. GENERALIZED RELATIVE INTEGRAL ACTION (RIA) METHOD

In this section, we present our main mathematical tool for analyzing the performance of DD sequences, which we refer to as the Relative Integral Action (RIA) method. To upper bound the distance between two unitary evolution operators, the RIA method introduces a new quantity that represents the integral of the difference between their generators. This idea has appeared in different context [1–3], and we adopt the term "integral action" from Ref.[3].

## A. First-order RIA theorem

Consider two unitary evolutions generated by a time-independent Hamiltonian $H_0$ and its perturbed version $H_1 = H_0 + V(t)$ with total evolution time $T$,

$$U_0(T) = e^{-iH_0T}, \quad U_1(T) = \mathcal{T}\exp\left(-i\int_0^T H_1 dt\right),$$ (S13)

where $\mathcal{T}$ denotes the time-ordering operator. Our objective is to bound the distance between $U_0(T)$ and $U_1(T)$. This distance can directly be upper-bounded by the distance between the two Hamiltonians $H_0$ and $H_1$. However, in highly *oscillatory* systems, the distance between Hamiltonians provides a loose bound. In such cases, a more appropriate quantity to consider is the difference between their integral actions [1–3]:

$$S(t) := \int_0^t (H_1 - H_0) dt_1 = \int_0^t V(t_1) dt_1. \qquad \text{(Relative integral action (RIA))}$$ (S14)

In general, we can bound the distance between the two unitary operations in terms of the RIA:

**Theorem S1** (RIA theorem). *Consider two unitary operations $U_0(T)$ and $U_1(T)$, generated by $H_0$ and $H_1 = H_0 + V(t)$ over the interval $t \in [0, T]$, respectively, and define $\beta = \|H_0\|$ and $J = \max_{t \in [0,T]} \|V(t)\|$. Let $S(t) = \int_0^t V(t_1) dt_1$ denote the RIA between the two Hamiltonians, with the condition $S(T) = 0$. Then, we have*

$$\|U_0(T) - U_1(T)\| \le T(2\beta + J) \cdot \max_{t \in [0,T]} \|S(t)\|.$$ (S15)

*Proof.* Since $\dot{U}_j(t) = -iH_j(t)U_j(t)$ for $j = 0, 1$, we have

$$\|U_0(T) - U_1(T)\| = \|U_0^\dagger(T)U_1(T) - I\| = \left\|\int_0^T (U_0^\dagger(t)U_1(t))' dt\right\| = \left\|\int_0^T U_0^\dagger(t)V(t)U_1(t) dt\right\|.$$ (S16)

Since $V(t) = S'(t)$, using integration by part with $S(T) = 0$, we have

$$\int_0^T U_0^\dagger(t)V(t)U_1(t) dt = U_0^\dagger(t)S(t)U_1(t)\Big|_0^T - i\int_0^T U_0^\dagger(t)(H_0 S(t) - S(t)H_0 - S(t)V(t))U_1(t) dt$$ (S17)

$$= -i\int_0^T U_0^\dagger(t)(H_0 S(t) - S(t)H_0 - S(t)V(t))U_1(t) dt,$$ (S18)

which gives the desired upperbound:

$$\|U_0(T) - U_1(T)\| \le T \cdot \max_{t \in [0,T]} \|H_0 S(t) - S(t)H_0 - S(t)V(t)\| \le T(2\beta + J) \cdot \max_{t \in [0,T]} \|S(t)\|.$$

$\square$

## B. Limitation of first-order RIA

Our goal is to analyze high-order DD methods. But the first-order version of the RIA fails to capture tight bounds for high-order DD protocols, which leads us to generalize the RIA method. Here we expand on this observation and motivate deriving a higher-order method.

Consider the system-bath Hamiltonian $H = H_S \otimes I_B + I_S \otimes H_B + H_{SB} + H_C(t) \otimes I_B$. In the interaction picture with respect to the evolution of $H_C(t)$ (which is also often phrased as a toggling frame), the new evolution is generated by a new Hamiltonian:

$$\widetilde{H} := H_S + H_B + \widetilde{H}_{SB}(t), \quad \widetilde{H}_{SB}(t) := U_C^\dagger(t)H_{SB}U_C(t), \quad U_C(t) = \mathcal{T}\exp\left(-i\int_0^t H_c(t_1) dt_1\right).$$ (S19)

Let $H_0 = H_S + H_B$. We want to compare the two unitary evolutions generated by the ideal evolution $H_0$ and the DD-protected evolution $H_1 = \widetilde{H} = H_0 + \widetilde{H}_{SB}(t)$. Hence, the RIA becomes $S(t) = \int_0^t V(t_1) dt_1$ where $V(t) = \widetilde{H}_{SB}(t)$.

As a simple example to first show the application of the first-order RIA method, consider the following interaction term:

$$H_{SB} = JZ_s \otimes B_z, \quad Z_s = \sum_{j=1}^{n} Z_j, \quad \|B_z\| = 1, \tag{S20}$$

and the DD protocol:

$$U_C(t) = (X^{\otimes n})^m, \quad t/\tau \in [m, m+1). \tag{S21}$$

Because $X^{\otimes n} Z_s X^{\otimes n} = -Z_s$, in the toggling frame, we have

$$\widetilde{H}_{SB}(t) = (-1)^m JZ_s \otimes B_z, \quad t/\tau \in [m, m+1). \tag{S22}$$

The RIA is the integral of $\widetilde{H}_{SB}(t)$, which becomes

$$S(t) = \begin{cases} (t - 2m\tau)JZ_s \otimes B_z, & t/\tau \in [2m, 2m+1), \\ [(2m+2)\tau - t]JZ_s \otimes B_z, & t/\tau \in [2m+1, 2m+2). \end{cases} \tag{S23}$$

Because $S(t)$ returns to 0 periodically, its norm is bounded by

$$\max_{t \in [0,T]} \|S(t)\| \leq J\tau. \tag{S24}$$

So, Theorem S1 gives

$$\|U_0(T) - U_1(T)\| \leq T(2\beta + J)J\tau, \tag{S25}$$

which captures the benefit of first order $DD$. However, this method fails to show the advantage of higher-order DD protocols such as concatenated DD (CDD) [4]. For instance, consider using the RIA method to analyze the performance of first-level CDD that uses $g_\ell = (I, X^{\otimes n}, X^{\otimes n}, I)$ pulse sequences on $H_{SB}$. The new $S(t)$ becomes

$$S(t) = \begin{cases} (t - 4m\tau)JZ_s \otimes B_z, & t/\tau \in [4m, 4m+1), \\ [(4m+2)\tau - t]JZ_s \otimes B_z, & t/\tau \in [4m+1, 4m+3), \\ [t - (4m+4)\tau]JZ_s \otimes B_z, & t/\tau \in [4m+3, 4m+4). \end{cases} \tag{S26}$$

Observe that $\max_{t \in [0,T]} \|S(t)\|$ is still $J\tau$. Thus, the bound derived from the RIA are the same for both non-concatenated and concatenated DD. However, the concatenated DD performs better than the first-order DD. To address this issue, we will generalize the original RIA method to provide tighter error bounds for CDD and other higher-order DD protocols.

### C. Higher-order RIA

Before presenting the main theorem for the higher-order RIA method, first, we motivate the approach using ideas similar to the Dyson series [5]. In the proof in Theorem S1, we applied integration by parts to obtain

$$U_0^\dagger(T)U_1(T) - I = (-i)^2 \int_0^T U_0^\dagger(t)([H_0, S(t)] - S(t)V(t))U_1(t)dt. \tag{S27}$$

If we treat $i[H_0, S(t)]$ as the new time-dependent perturbation, we can apply the integration by parts procedure once more, yielding

$$U_0^\dagger(T)U_1(T) - I = (-i)^2 \int_0^T U_0^\dagger(t)(S^{(2)}(t))'U_1(t)dt - (-i)^2 \int_0^T U_0^\dagger(t)S(t)V(t)U_1(t)dt, \tag{S28}$$

where we defined

$$S^{(2)}(t) := \int_0^t [H_0, S(t_1)] \, dt_1 = \left[ H_0, \int_0^t dt_1 \int_0^{t_1} dt_2 V(t_2) \right]. \tag{S29}$$

A second-order DD sequence suppresses the perturbation term up to second-order in $T$ and we have $S^{(2)}(T) = 0$. Using integration by parts on the first term in the RHS of Eq.(S28) yields:

$$\|U_0(T) - U_1(T)\| \leq T(2\beta + J) \cdot \max_{t \in [0,T]} \|S^{(2)}(t)\| + \mathcal{O}(J^2). \tag{S30}$$

The bound $\max_{t \in [0,T]} \|S^{(2)}(t)\| = \mathcal{O}(\beta J \tau^2)$, improves the bound by first-order RIA method with a factor of $\tau$, and matches to the expected error bound of a second order DD protocol.

Similarly, we can further define $S^{(3)}(t) := \int_0^t [H_0, S^{(2)}(t')]dt'$, and so on. By recursion, we define

$$S^{(K)}(t) := \text{ad}_{H_0}^{K-1}\left[\int_0^t dt_1 \int_0^{t_1} dt_2 \cdots \int_0^{t_{K-1}} dt_K V(t_K)\right], \qquad \text{(Generalized RIA)} \tag{S31}$$

where $\text{ad}_O(\cdot) := [O, \cdot]$. When the DD sequences ensure that $S^{(k)}(T) = 0$ is satisfied, we can apply additional steps of integration by parts, leading to an improved upper bound for $\|U_0(T) - U_1(T)\|$. Such condition can be formally defined as follows:

**Definition S1** ($K$-th order integral condition). *Given a time-dependent operator $V(t)$ where $t \in [0,T]$, if $\exists K \in \mathbb{Z}$ such that for all integer in $1 \leq k \leq K$, we have*

$$\int_0^T dt_1 \int_0^{t_1} dt_2 \cdots \int_0^{t_k-1} dt_k V(t_k) = 0, \tag{S32}$$

*then we say $V(t)$ satisfies the $K$-th order integral condition.*

It is clear that if $V(t)$ satisfies the $K$-th order integral condition, then $S^{(K)}(T) = 0$. With this definition, we can proceed to present the main theorem of this section:

**Theorem S2** (Generalized RIA theorem). *Consider a time-independent Hamiltonian $H_0$ and a time-dependent perturbation $V(t)$. Let $U_0$ be the unitary evolution generated by $H_0$ with $U_0(0) = I$, and let $U_1$ be generated by $H_0 + V(t)$ with $U_1(0) = I$. Define $\beta = \|H_0\|$ and $J = \max_{t \in [0,T]} \|V(t)\|$.*

*If the perturbation $V(t)$ satisfies the $K$-th order integral condition (as defined in Definition S1), then $\|U_1(T) - U_0(T)\|$ is upper bounded by*

$$T(2\beta)^K \max_{t \in [0,T]} \left\|\int_0^t dt_1 \cdots \int_0^{t_{K-1}} dt_K V(t_K)\right\| + T \max_{t \in [0,T]} \|V(t)S(t)\| + \frac{T^2}{2} J(2\beta + J) \max_{t \in [0,T]} \|S(t)\|. \tag{S33}$$

*Proof.* For every $t$, we have

$$U_0^\dagger(t)U_1(t) = I - i\int_0^t U_0^\dagger(t_1)V(t_1)U_1(t_1)dt_1$$
$$= I - i\int_0^t U_0^\dagger(t_1)V(t_1)U_0(t_1)dt_1 - \int_0^t dt_1 U_0^\dagger(t_1)V(t_1)U_0(t_1)\int_0^{t_1} dt_2 U_0^\dagger(t_2)V(t_2)U_1(t_2). \tag{S34}$$

where the second equality is obtained by replacing $U_1(t_1)$ in the first integral with $U_0(t_1) - iU_0(t_1)\int_0^{t_1} U_0^\dagger(t_2)V(t_2)U_1(t_2)dt_2$. By setting $t = T$ and applying integration by parts recursively, the second term in the RHS of Eq.(S34) can be expressed as

$$-i\int_0^T U_0^\dagger(t)V(t)U_0(t)dt = -iU_0^\dagger(t)S(t)U_0(t)|_0^T + (-i)^2\int_0^T U_0^\dagger(t)\underbrace{[H_0, S(t)]}_{(S^{(2)}(t))'}U_0(t)dt \tag{S35}$$

$$= U_0^\dagger(t)[-iS(t) + (-i)^2 S^{(2)}(t)]U_0(t)|_0^T + (-i)^3\int_0^T U_0^\dagger(t)\underbrace{[H_0, S^{(2)}(t)]}_{(S^{(3)}(t))'}U_0(t)dt = \ldots \tag{S36}$$

$$= \sum_{k=1}^K (-i)^k U_0^\dagger(t)S^{(k)}(t)U_0(t)|_0^T + (-i)^K\int_0^T U_0^\dagger(t)[H_0, S^{(K)}(t)]U_0(t)dt. \tag{S37}$$

where the generalized RIA $S^{(k)}$ is defined in Eq.(S31). Since we assume that $V(t)$ satisfies the $K$-th order integral condition, we have $S^{(k)} = 0$, $1 \leq \forall k \leq K$, so that the first term in Eq.(S37) vanishes. Therefore, we have

$$\left\| -i \int_0^T U_0^\dagger(t) V(t) U_0(t) dt \right\| = \left\| (-i)^K \int_0^T U_0^\dagger(t) [H_0, S^{(K)}(t)] U_0(t) \right\| \tag{S38}$$

$$\leq T(2\beta) \cdot \max_{t \in [0,T]} \|S^{(K)}(t)\| \leq T(2\beta)^K \cdot \max_{t \in [0,T]} \left\| \int_0^t dt_1 \cdots \int_0^{t_{K-1}} dt_K V(t_K) \right\|. \tag{S39}$$

Next, let us bound the last term in the RHS of Eq.(S34). First, note that the integration by part gives

$$\int_0^{t_1} U_0^\dagger(t_2) V(t_2) U_1(t_2) dt_2 = U_0^\dagger(t_1) S(t_1) U_1(t_1) - i \int_0^{t_1} U_0^\dagger(t_2)(H_0 S(t_2) - S(t_2) H_0 - S(t_2) V(t_2)) U_1(t_2) dt_2, \tag{S40}$$

which gives the upper bound of

$$\left\| \int_0^T dt_1 U_0^\dagger(t_1) V(t_1) U_0(t_1) \int_0^{t_1} dt_2 U_0^\dagger(t_2) V(t_2) U_1(t_2) \right\| \tag{S41}$$

$$\leq \left\| \int_0^T dt U_0^\dagger(t) V(t) S(t) U_1(t) \right\| + \left\| \int_0^T dt_1 U_0^\dagger(t_1) V(t_1) U_0(t_1) \int_0^{t_1} dt_2 U_0^\dagger(t_2)(H_0 S(t_2) - S(t_2) H_0 - S(t_2) V(t_2)) U_1(t_2) \right\|. \tag{S42}$$

The first term in Eq.(S42) is upper bounded by

$$\left\| \int_0^T dt U_0^\dagger(t) V(t) S(t) U_1(t) \right\| \leq T \max_{t \in [0,T]} \|V(t) S(t)\|, \tag{S43}$$

and the second term in Eq.(S42) is upper bounded by $\frac{T^2}{2} J(2\beta + J) \cdot \max_{t \in [0,T]} \|S(t)\|$. Hence, Eq.(S42) is upper bounded by

$$\left\| \int_0^T dt_1 U_0^\dagger(t_1) V(t_1) U_0(t_1) \int_0^{t_1} dt_2 U_0^\dagger(t_2) V(t_2) U_1(t_2) \right\| \leq T \max_{t \in [0,T]} \|V(t) S(t)\| + \frac{T^2}{2} J(2\beta + J) \cdot \max_{t \in [0,T]} \|S(t)\|. \tag{S44}$$

Combining Eqs. (S34), (S39) and (S44) completes the proof. □

Since $\|S(t)\| \leq JT$, the above theorem states that $\|U_1(T) - U_0(T)\| = \mathcal{O}(JT^{K+1}) + \mathcal{O}(J^2 T^2)$. In fact, under certain conditions on $S(t)$ and $V(t)$, we can obtain a tighter bound of $\|U_1(T) - U_0(T)\| = \mathcal{O}(JT^{K+1}) + \mathcal{O}(J^2 T^3)$:

**Proposition S1.** *Consider a time-independent Hamiltonian $H_0$ and a time-dependent perturbation $V(t)$. Let $U_0$ be the unitary evolution generated by $H_0$ with $U_0(0) = I$, and let $U_1$ be generated by $H_0 + V(t)$ with $U_1(0) = I$. Define $\beta = \|H_0\|$ and $J = \max_{t \in [0,T]} \|V(t)\|$.*

*If the perturbation $V(t)$ satisfies the $K$-th order integral condition (as defined in Definition (S1)) and the following condition:*

$$\int_0^T dt V(t) S(t) = 0. \tag{S45}$$

*then $\|U_1(T) - U_0(T)\|$ is upper bounded by*

$$T(2\beta)^K \max_{t \in [0,T]} \left\| \int_0^t dt_1 \cdots \int_0^{t_{K-1}} dt_K V(t_K) \right\| + T(2\beta + J) \max_{t \in [0,T]} \left\| \int_0^t dt_1 V(t_1) S(t_1) \right\| + \frac{T^2}{2} J(2\beta + J) \max_{t \in [0,T]} \|S(t)\|. \tag{S46}$$

*Proof.* The proof of this proposition follows the same steps as the proof of Theorem S2 except for the upper bound of the first term in Eq.(S42). Instead of using the triangle inequality as before, we use the integration by parts to obtain

$$\int_0^T dt U_0^\dagger(t) V(t) S(t) U_1(t) = U_0^\dagger(t) \left( \int_0^t V(t_1) S(t_1) dt_1 \right) U_1(t) \Big|_0^T \tag{S47}$$

$$- i \int_0^T U_0^\dagger \left[ H_0, \left( \int_0^t V(t_1) S(t_1) dt_1 \right) \right] U_1(t) dt - i \int_0^T U_0^\dagger \left( \int_0^t V(t_1) S(t_1) dt_1 V(t) \right) U_1(t) dt. \tag{S48}$$

Due to our assumption on Eq.(S45), the first term in the RHS of the above equation vanishes. Hence, we obtain a tighter upper bound:

$$\left\| \int_0^T dt\, U_0^\dagger(t) V(t) S(t) U_1(t) \right\| \leq T(2\beta + J) \cdot \max_{t \in [0,T]} \left\| \int_0^t dt_1 V(t_1) S(t_1) \right\|. \tag{S49}$$

$\square$

In general, the condition $\int_0^T V(t)S(t)dt = 0$ does not hold, but there are two sufficient conditions where it does. The first is when $\forall t, [V(t), S(t)] = 0$, because the integration by parts gives $\int_0^T V(t)S(t)dt = -\int_0^T S(t)V(t)dt$. The second case is when $V(t)$ exhibits time-reversal symmetry, i.e., $\forall t, V(t) = V(T-t)$. In this case, the following lemma guarantees that the condition is satisfied:

**Proposition S2.** *Suppose $V(t)$ satisfies $S(T) = 0, V(t) = V(T-t)$. Then*

$$\int_0^T V(t)S(t)dt = 0. \tag{S50}$$

*Proof.* Because $V(t) = V(T-t)$, we have $\int_0^{T/2} V(t)dt = \int_0^{T/2} V(T-t)dt = \int_T^{T/2} V(t)d(-t) = \int_{T/2}^T V(t)dt$, and the condition $S(T) = \int_0^T V(t)dt = 0$ ensures that

$$\int_0^{T/2} V(t)dt = \int_{T/2}^T V(t)dt = 0. \tag{S51}$$

Hence, for all $t \in [0, T/2]$, we have

$$S(t) = -\int_t^{T/2} V(t')dt', \tag{S52}$$

$$S(T-t) = \int_0^{T-t} V(t')dt' = \int_{T/2}^{T-t} V(t')dt' = \int_{T/2}^t V(T-t')d(-t') = -S(t). \tag{S53}$$

Finally,

$$\int_0^T V(t)S(t)dt = \int_0^{T/2} V(t)S(t)dt + \int_{T/2}^T V(t)S(t)dt = \int_0^{T/2} V(t)S(t)dt + \int_0^{T/2} V(t)S(T-t)dt = 0. \tag{S54}$$

$\square$

### D. Example 1: Application of higher-order RIA to bound symmetric universal decoupling (XY8)

In this section, we utilize our generalized RIA method to rigorously analyze the performance of symmetric universal decoupling, also known as XY8. Using the notation introduced in the main text, let $G = \{g\}$ represent the decoupling group of the XY8 sequence, which has the equal pulse interval $\tau = \tau_\ell \ \forall \ell$.

The XY8 sequence is a time-symmetric variant of the XY4 sequence. Specifically, the pulse sequences are described as

$$\text{XY4}: \quad f_\tau Y f_\tau X f_\tau Y f_\tau X, \tag{S55}$$

$$\text{XY8}: \quad X f_\tau Y f_\tau X f_\tau Y f_\tau f_\tau Y f_\tau X f_\tau Y f_\tau X, \tag{S56}$$

where $f_\tau = e^{-iH\tau}$ represents the free evolution unitary operator. Note that replacing either $X$ or $Y$ pulses with $Z$ pulses does not affect the performance and is still considered as the XY4 or XY8 sequences.

In the interaction picture with respect to the evolution of control Hamiltonian $H_C(t)$, the system-bath interaction term $H_{SB}$ becomes:

$$V(t) = g_\ell H_{SB} g_\ell^\dagger, \quad \ell\tau \leq t < (\ell+1)\tau. \tag{S57}$$

In this section, we focus on the 1-local bath $H_{SB} = X \otimes B_X + Y \otimes B_Y + Z \otimes B_Z$, but the result can be generalized to other situations. For XY8, in the notation of Theorem 1, we have

$$G = \{g_0, g_2, g_2, g_3\} = \{I^{\otimes n}, X^{\otimes n}, Y^{\otimes n}, Z^{\otimes n}\}, \quad L = 8, \quad \tau_\ell = \tau, \quad T = 8\tau. \tag{S58}$$

**Theorem S3** (Rigorous performance bound on XY8). *Consider the Hamiltonian $H = H_0 + H_{SB}$ where $H_{SB} = X \otimes B_X + Y \otimes B_Y + Z \otimes B_Z$, and $J = \|B_X\| + \|B_Y\| + \|B_Z\|$, $\beta = \|H_0\|$. The XY8 protocol yields the error bound:*

$$\|D - U_0\| \le 128 J \beta^2 \tau^3 + 96(2\beta + J) J^2 \tau^3. \tag{S59}$$

*Proof.* One can verify that in the toggling frame

$$V(t) = x_\ell X \otimes B_X + y_\ell Y \otimes B_Y + z_\ell Z \otimes B_Z, \quad \ell\tau \le t < (\ell+1)\tau, \tag{S60}$$

$$(x_\ell)_{\ell=0}^7 = (1, 1, -1, -1, -1, -1, 1, 1), \tag{S61}$$

$$(y_\ell)_{\ell=0}^7 = (1, -1, 1, -1, -1, 1, -1, 1), \tag{S62}$$

$$(z_\ell)_{\ell=0}^7 = (1, -1, -1, 1, 1, -1, -1, 1). \tag{S63}$$

Note that $V(t)$ satisfies the second order integral condition and $V(t) = V(T-t)$. Hence we can use Proposition S1 with $K = 2$:

$$\|D - U_0\| \le T(2\beta)^2 \max_{t \in [0,T]} \left\| \int_0^t dt_1 \int_0^{t_1} dt_2 V(t_2) \right\| + T(2\beta + J) \max_{t \in [0,T]} \left\| \int_0^t dt_1 V(t_1) S(t_1) \right\| + \frac{T^2}{2} J(2\beta + J) \max_{t \in [0,T]} \|S(t)\|. \tag{S64}$$

After computation, we obtain

$$\max_{t \in [0,T]} \left\| \int_0^t dt_1 \int_0^{t_1} dt_2 V(t_2) \right\| \le 4\tau^2 \|B_X\| + 2\tau^2 \|B_Y\| + \tau^2 \|B_Z\| \le 4J\tau^2, \tag{S65}$$

$$\max_{t \in [0,T]} \|S(t)\| \le \tau(\|B_X\| + \|B_Y\| + \|B_Z\|) \le J\tau, \quad \max_{t \in [0,T]} \left\| \int_0^t dt_1 V(t_1) S(t_1) \right\| \le J^2 T\tau. \tag{S66}$$

Therefore,

$$\|D - U_0\| \le 128 J \beta^2 \tau^3 + 96(2\beta + J) J^2 \tau^3. \tag{S67}$$

$\square$

### E. Example 2: Application of higher-order RIA to bound CDD performance

In this section, we utilize our generalized RIA method to rigorously analyze the performance of concatenated DD (CDD) [4], described in Eq.(11). Using the notation introduced in the main text, let $G = \{g\}$ represent the decoupling group of a given CDD protocol, which has the equal pulse interval $\tau = \tau_\ell$ $\forall \ell$. In the interaction picture with respect to the evolution of control Hamiltonian $H_C(t)$, the system-bath interaction term $H_{SB}$ becomes:

$$V(t) = g_\ell H_{SB} g_\ell^\dagger, \quad \ell\tau \le t < (\ell+1)\tau. \tag{S68}$$

In this section, we focus on the 1-local bath $H_{SB} = X \otimes B_X + Y \otimes B_Y + Z \otimes B_Z$ and concatenated universal decoupling together with its randomized version as instances, but the result can be generalized to other situations.

For the $\text{CDD}_K$, in the notation of Theorem 1, we have

$$G = \{g_0, g_2, g_2, g_3\} = \{I^{\otimes n}, X^{\otimes n}, Y^{\otimes n}, Z^{\otimes n}\}, \quad L = 4^K, \quad \tau_\ell = \tau. \tag{S69}$$

Note that $n$ is divisible by 4 so that $G$ forms a group. Suppose in the digit-4 representation, $\ell = \ell_0 + 4\ell_1 + \cdots + 4^{K-1}\ell_{K-1}$ where $\ell_k \in \{0, 1, 2, 3\}$. Then

$$g_\ell = g_{\ell_{K-1}} g_{\ell_{K-2}} \cdots g_{\ell_1} g_{\ell_0} g_{\ell_1}^\dagger \cdots g_{\ell_{K-2}}^\dagger g_{\ell_{K-1}}^\dagger, \quad \ell = 0, 1, \cdots, 4^K - 1. \tag{S70}$$

**Theorem S4** (Rigorous performance bound on CDD; Derivation of Eq.(11).). *Consider the Hamiltonian $H = H_0 + H_{SB}$ where $H_{SB} = X \otimes B_X + Y \otimes B_Y + Z \otimes B_Z$, and $J = \|B_X\| + \|B_Y\| + \|B_Z\|$, $\beta = \|H_0\|$. The deterministic $\text{CDD}_K$ protocol yields the error bound:*

$$\|D - U_0\| \le T(2\beta)^K J 2^{K(K-1)+1} \tau^K + \left[ 2JT + JT^2(2\beta + J) \right] J\tau = \mathcal{O}(J\beta^K \tau^{K+1}) + \mathcal{O}(J^2\tau^2). \tag{S71}$$

*Proof.* One can verify that in the toggling frame

$$V(t) = f_X(t)X \otimes B_X + f_Y(t)Y \otimes B_Y + f_Z(t)Z \otimes B_Z. \tag{S72}$$

with $f_X(t), f_Y(t), f_Z(t)$ defined in Proposition S3. In the same proposition, we prove that $V(t)$ satisfies the $K$-th order integral condition, and

$$\max_{t \in [0,T]} \left\| \int_0^t dt_1 \cdots \int_0^{t_{K-1}} dt_K V(t_K) \right\| \le \|B_X\| \cdot 2^{K(K-1)+1} \tau^K + \|B_Y\| \cdot 2^{(K-1)^2} \tau^K + \|B_Z\| \cdot 2^{(2K-1)(K-1)} \tau^{2K}. \tag{S73}$$

Recall that $J = \|B_X\| + \|B_Y\| + \|B_Z\|$. In the worst case where $\|B_Y\| = \|B_Z\| = 0$, the upper bound becomes $J 2^{K(K-1)+1} \tau^K$. By virtue of Eq. (S33) in Theorem S2 and the fact that $\|S(t)\| \le 2J\tau$, the error bound of the deterministic $\mathrm{CDD}_K$ protocol can be obtained. $\qquad \square$

Considering $\mathcal{O}(J)$ terms, our result has an improvement factor of order $(4 \cdot 1.027)^K$ compared to results obtained by Magnus expansion in Ref.[6]. Additionally, our derivation accounts for all higher-order contributions.

Before presenting the key result that is used to prove the above theorem, the following lemma is useful in the analysis of $V(t)$.

**Lemma S1.** *Let $(y_0, y_1, \cdots, y_{m-1})$ be an array of real numbers such that $\sum_{i=1}^m y_i = 0, y_0 = 1$. For all $n \in \mathbb{N}$, suppose it has digit-m representation $n = n_0 + mn_1 + m^2 n_2 + \cdots$, and let $x_n = \prod_k y_{n_k}$. Then for any polynomial $p(n)$ with order $k_1$, we have*

$$\sum_{n=0}^{m^{k_2}-1} x_n p(n) = 0 \quad \forall k_2 > k_1. \tag{S74}$$

*Proof.* Note that

$$\begin{aligned}
\sum_{n=0}^{m^{k_2}-1} x_n p(n) &= \sum_{n'=0}^{m^{k_2-1}-1} x_{n'} \left[ \sum_{k=0}^{m-1} y_k p(mn' + k - 1) \right] \\
&= \sum_{n'=0}^{m^{k_2-1}-1} x_{n'} p'(n')
\end{aligned} \tag{S75}$$

The order of $p'$ is a polynomial of $n'$ with order $k_1 - 1$. Using this relation recursively, the lemma is proved. $\qquad \square$

We now present our key lemma used in the proof of Theorem S4.

**Proposition S3.** *On region $t \in [0, T]$, $T = 4^K \tau$, define*

$$f_X(t) = \prod_{k=0}^{K-1} x_{\ell_k}, \quad \{x_\ell\}_{\ell=0}^3 = (1, 1, -1, -1), \quad \lfloor t/\tau \rfloor = \ell; \tag{S76}$$

$$f_Y(t) = \prod_{k=0}^{K-1} y_{\ell_k}, \quad \{y_\ell\}_{\ell=0}^3 = (1, -1, 1, -1), \quad \lfloor t/\tau \rfloor = \ell; \tag{S77}$$

$$f_Z(t) = \prod_{k=0}^{K-1} z_{\ell_k}, \quad \{z_\ell\}_{\ell=0}^3 = (1, -1, -1, 1), \quad \lfloor t/\tau \rfloor = \ell. \tag{S78}$$

*For all $1 \le k \le K$, $\alpha = X, Y, Z$, we have*

$$\int_0^T dt_1 \int_0^{t_1} dt_2 \cdots \int_0^{dt_{K-1}} dt_K f_\alpha(t_K) = 0. \tag{S79}$$

*The $K$-th order integrals of the functions have maximal values:*

$$\max_t \left| \int_0^t dt_1 \int_0^{dt_1} dt_2 \cdots \int_0^{dt_{K-1}} dt_K f_X(t) \right| = 2^{K(K-1)+1} \tau^K, \tag{S80}$$

$$\max_t \left| \int_0^t dt_1 \int_0^{dt_1} dt_2 \cdots \int_0^{dt_{K-1}} dt_K f_Z(t) \right| = 2^{(2K-1)(K-1)} \tau^{2K}, \tag{S81}$$

$$\max_t \left| \int_0^t dt_1 \int_0^{dt_1} dt_2 \cdots \int_0^{dt_{K-1}} dt_K f_Y(t) \right| = 2^{(K-1)^2} \tau^K, \tag{S82}$$

$$\tag{S83}$$

*Proof.* We first prove the bound Eq.(S79). Observe that

$$\int_0^T dt_1 \int_0^{t_1} dt_2 \cdots \int_0^{t_{K-1}} dt_K f_Z(t_K) = \int_0^T dt_K \int_{t_K}^T dt_{K-1} \cdots \int_{t_2}^T dt_1 f_Z(t_K) \tag{S84}$$

$$= \frac{1}{(K-1)!} \int_0^T f_Z(t)(T-t)^{K-1} dt. \tag{S85}$$

Hence,

$$\int_0^T f_Z(t)(T-t)^{K-1} dt = \sum_{\ell=0}^{4^K-1} \prod_{k=1}^K z_{\ell_k} \int_{\ell\tau}^{(\ell+1)\tau} (T-t)^{K-1} dt \tag{S86}$$

$$= \sum_{\ell=0}^{4^K-1} \prod_{k=0}^{K-1} z_{\ell_k} \frac{\tau^K}{K!} \left[ (4^K - \ell - 1)^K - (4^K - \ell)^K \right] \tag{S87}$$

$$= \frac{\tau^K}{K!} \sum_{\ell=0}^{4^K-1} \prod_{k=0}^{K-1} z_{\ell_k} p_\ell, \quad (p_\ell \text{ is a polynomial of } \ell \text{ with order } K-1). \tag{S88}$$

$$= 0 \quad \text{(Lemma S1)}. \tag{S89}$$

The proof for $f_X(t), f_Y(t)$ is the same.

In order to compute the upper bound, we start with the following case. Given $K$, suppose $\lfloor t/\tau \rfloor = \ell, \ell = \ell_0 + 2\ell_1 + \cdots + 2^{K-1}\ell_{K-1}, \ell_k \in \{0, 1\}$. Then let

$$f(t) = \prod_{k=0}^{K-1} (-1)^{\ell_k}. \tag{S90}$$

Define

$$F(t; 1) = \begin{cases} 1, & 0 \le t < \tau, \\ -1, & \tau \le t < 2\tau. \end{cases}, \quad F(t; k+1) = \begin{cases} \int_0^t dt' F(t'; k), & 0 \le t < 2^k\tau, \\ -F(t - 2^k\tau; k+1), & 2^k\tau \le t < 2^{k+1}\tau. \end{cases} \tag{S91}$$

Observe that $\int_0^t dt_1 \cdots \int_0^{dt_{K-2}} dt_{K-1} f(t_{K-1}) = F(t; K)$. Using mathematical induction, one can prove that for all $1 \le k \le K$, the function $F(t; k)$ satisfies

1. $F(t; k) > 0, \quad 0 < t < 2^{k-1}\tau$.

2. $F(t; k) < 0, \quad 2^{k-1}\tau < t < 2^k\tau$.

3. $F(2^{k-1}\tau; k) = 0$, which is also the only zero point of the function other than $t = 0, 2^k\tau$.

Suppose $K \ge 2$. Because the second half of $F(t; K)$ is simply the inverse of the first half, we have

$$\max_{t \in [0, 2^K\tau]} |F(t; K)| = \max_{t \in [0, 2^{K-1}\tau]} |F(t; K)| = \max_{t \in [0, 2^{K-1}\tau]} \int_0^t dt' F(t'; K-1). \tag{S92}$$

Since $F(t; K - 1) = 0$ only occurs at $t = 0, 2^{K-2}\tau, 2^{K-1}\tau$, the maximal value of $F(t; K)$ must be taken at $t = 2^{K-2}\tau$. Hence,

$$\max_{t \in [0, 2^K \tau]} |F(t; K)| = F(2^{K-2}\tau; K). \tag{S93}$$

Finally,

$$F(2^{K-1}\tau; K + 1) = \int_0^{2^{K-1}\tau} dt_1 \int_0^{dt_1} dt_2 \cdots \int_0^{dt_{K-1}} dt_K f(t) \tag{S94}$$

$$= \frac{1}{(K-1)!} \int_0^{2^{K-1}\tau} f(t)(2^{K-1}\tau - t)^{K-1} dt \tag{S95}$$

$$= \frac{\tau^K}{(K-1)!} \sum_{\ell=0}^{2^{K-1}-1} \prod_{k=0}^{K-1} (-1)^{\ell_k} \int_\ell^{\ell+1} (2^{K-1} - x)^{K-1} dx \qquad (\ell_K = 1) \tag{S96}$$

$$= \frac{\tau^K}{K!} \sum_{\ell=0}^{2^{K-1}-1} \prod_{k=0}^{K-1} (-1)^{\ell_k} \left[ (2^{K-1} - \ell - 1)^K - (2^{K-1} - \ell)^K \right] \tag{S97}$$

$$= \frac{\tau^K}{(K-1)!} \sum_{\ell=0}^{2^{K-1}-1} \prod_{k=0}^{K-1} (-1)^{\ell_k} (-\ell)^{K-1}(-1) \qquad \text{(Lemma S1)} \tag{S98}$$

$$= \frac{\tau^K}{(K-1)!} \sum_{\ell=0}^{2^{K-2}-1} \prod_{k=0}^{K-2} (-1)^{\ell_k} [(-2\ell - 1)^{K-1} - (-2\ell)^{K-1}] \tag{S99}$$

$$= \frac{\tau^K}{(K-1)!} \sum_{\ell=0}^{2^{K-2}-1} \prod_{k=0}^{K-2} (-1)^{\ell_k} [(-1)(K-1)(-2\ell)^{K-2}] \qquad \text{(Lemma S1)} \tag{S100}$$

$$= \frac{\tau^K}{(K-1)!}(K-1)2^{K-2} \sum_{\ell=0}^{2^{K-2}-1} \prod_{k=0}^{K-2} (-1)^{\ell_k} (-\ell)^{K-2}(-1) = \cdots \tag{S101}$$

$$= 2^{(K-1)(K-2)/2}\tau^K. \tag{S102}$$

Thus,

$$\max_{t \in [0, 2^K \tau]} \left| \int_0^t dt_1 \int_0^{dt_1} dt_2 \cdots \int_0^{dt_{K-1}} dt_K f(t) \right| = 2^{(K-1)(K-2)/2}\tau^K. \tag{S103}$$

Note that the computation result of $f_Z(t)$ can be obtained from this example by setting $K \to 2K$. That is

$$\max_{t \in [0, 4^K \tau]} \left| \int_0^t dt_1 \int_0^{dt_1} dt_2 \cdots \int_0^{dt_{K-1}} dt_K f_Z(t) \right| = 2^{(2K-1)(K-1)}\tau^{2K}. \tag{S104}$$

As to $f_X(t)$, using the same argument, we can prove that the maximal value of its $K$-th order integral is achieved at $t = 2 \cdot 4^{K-1}\tau$, and

$$\int_0^{2 \cdot 4^{K-1}\tau} dt_1 \int_0^{dt_1} dt_2 \cdots \int_0^{dt_{K-1}} dt_K f_X(t) \tag{S105}$$

$$= \frac{1}{(K-1)!} \int_0^{2 \cdot 4^{K-1}\tau} f_X(t)(2 \cdot 4^{K-1}\tau - t)^{K-1} dt \tag{S106}$$

$$= \frac{1}{(K-1)!} \int_0^{4^{K-1}\tau} f_X(t)(2 \cdot 4^{K-1}\tau - t)^{K-1} dt + \frac{1}{(K-1)!} \int_0^{4^{K-1}\tau} f_X(t)(4^{K-1}\tau - t)^{K-1} dt \tag{S107}$$

The second equality holds because $f_X(t) = f_X(t + 4^{K-1}\tau)$.

$$\frac{1}{(K-1)!} \int_0^{4^{K-1}\tau} f_X(t)(4^{K-1}\tau - t)^{K-1} dt \tag{S108}$$

$$= \frac{\tau^K}{K!} \sum_{\ell=0}^{4^{K-1}-1} \prod_{k=0}^{K-1} x_{\ell_k} \left[ (4^{K-1} - \ell - 1)^K - (4^{K-1} - \ell)^K \right] \tag{S109}$$

$$= \frac{\tau^K}{(K-1)!} \sum_{\ell=0}^{4^{K-1}-1} \prod_{k=0}^{K-1} x_{\ell_k} (-\ell)^{K-1}(-1) \qquad \text{(Lemma S1)} \tag{S110}$$

$$= \frac{\tau^K}{(K-1)!} \sum_{\ell=0}^{4^{K-2}-1} \prod_{k=0}^{K-2} x_{\ell_k} (-1)^{K-1}[(4\ell + 3)^{K-1} + (4\ell + 2)^{K-1} - (4\ell + 1)^{K-1} - (4\ell)^{K-1}] \tag{S111}$$

$$= \frac{\tau^K}{(K-1)!} \sum_{\ell=0}^{4^{K-2}-1} \prod_{k=0}^{K-2} x_{\ell_k} (-1)^{K-1} \cdot 4 \cdot (K-1) \cdot (4\ell)^{K-2} \qquad \text{(Lemma S1)} \tag{S112}$$

$$= \frac{\tau^K}{(K-1)!} (K-1) 4^{K-1} \sum_{\ell=0}^{4^{K-2}-1} \prod_{k=0}^{K-2} x_{\ell_k} (-\ell)^{K-2}(-1) = \cdots \tag{S113}$$

$$= 2^{K(K-1)} \tau^K. \tag{S114}$$

The other term in Eq. (S107) has the same value. Thus,

$$\max_t \left| \int_0^t dt_1 \int_0^{t_1} dt_2 \cdots \int_0^{t_{K-1}} dt_K f_X(t) \right| = 2^{K(K-1)+1} \tau^K, \tag{S115}$$

The proof for $f_Y(t)$ is more tricky. We use $f_Y(t)$ as an instance. Given $K$, suppose $\lfloor t/\tau \rfloor = \ell, \ell = \ell_0 + 4\ell_1 + \cdots + 4^{K-1}\ell_{K-1}$. Then

$$f_Y(t) = \prod_{k=0}^{K-1} (-1)^{\ell_k}. \tag{S116}$$

Define

$$F_Y(t; 1) = \begin{cases} 1, & 0 \leq t < \tau, \\ -1, & \tau \leq t < 2\tau, \\ 1, & 2\tau \leq t < 3\tau, \\ -1, & 3\tau \leq t < 4\tau. \end{cases}, \quad F_Y(t; k+1) = \begin{cases} \int_0^t dt' F_Y(t'; k), & 0 \leq t < 4^k\tau, \\ -F_Y(t - 4^k\tau; k+1), & 4^k\tau \leq t < 2 \cdot 4^k\tau, \\ F_Y(t - 2 \cdot 4^k\tau; k+1), & 2 \cdot 4^k\tau \leq t < 3 \cdot 4^k\tau, \\ -F_Y(t - 3 \cdot 4^k\tau; k+1), & 3 \cdot 4^k\tau \leq t < 4^{k+1}\tau. \end{cases} \tag{S117}$$

Observe that $\int_0^t dt_1 \cdots \int_0^{dt_{K-2}} dt_{K-1} f_Y(t_{K-1}) = F_Y(t; K)$. Using mathematical induction, one can prove that for all $1 \leq k \leq K-1$, the function $F_Y(t; k)$ satisfies

1. $F_Y(t; k+1) \geq 0, \quad t \in (0, 4^k\tau) \cup (2 \cdot 4^k\tau, 3 \cdot 4^k\tau)$.

2. $F_Y(t; k+1) \leq 0, \quad t \in (4^k\tau, 2 \cdot 4^k\tau) \cup (3 \cdot 4^k\tau, 4^{k+1}\tau)$.

In other words, the derivative of $F_Y(t; k)$ is always $\geq 0, \leq 0, \geq 0, \leq 0$ in four regions arranged in order. Based on these observations, the maximal value of $F_Y(t; K)$ must be taken at $t = 4^{K-1}\tau$, which is the end of the first region. After a similar computation, we obtain

$$\max_t \left| \int_0^t dt_1 \int_0^{dt_1} dt_2 \cdots \int_0^{dt_{K-1}} dt_K f_Y(t) \right| = 2^{(K-1)^2} \tau^K, \tag{S118}$$

$\square$

## III. PROOF OF THEOREM 1

Our proposed sequence-randomized formula relies on applying unitary operations randomly to the given deterministic DD protocol. The following mixing lemma is a key tool for analyzing its performance.

**Lemma S2** (Mixing lemma [7, 8]). *Suppose $\{U_n\}_{n=0}^{N-1}$ is a set of unitaries, $\{p_n\}_{n=0}^{N-1}$ is a probability distribution. Let*

$$\mathcal{C}_1(X) = \sum_{n=0}^{N-1} p_n U_n X U_n^\dagger, \quad \mathcal{C}_2(X) = V X V^\dagger \tag{S119}$$

*where $V$ is unitary. Then*

$$\|\mathcal{C}_1 - \mathcal{C}_2\|_\diamond \leq \max_n \|U_n - V\|^2 + 2 \left\| \sum_{n=0}^{N-1} p_n U_n - V \right\|. \tag{S120}$$

Then, we have our main result:

**Theorem S5** (Randomized DD error bound; Restatement of Theorem 1). *Let $H_S$ be the system Hamiltonian, $H_B$ the bath Hamiltonian and $H_{SB}$ the interaction between systems and bath. Denote the ideal (decoupled) evolution by $U_0 = \exp(-iH_0 T)$ where $H_0 = H_S + H_B$ and $T = \tau L$. Define $J = \|H_{SB}\|$ and $\beta = \|H_0\|$. Let $g_\ell \in G$ be the elements of the decoupling group satisfying $\forall \ell$, $[g_\ell, H_0] = 0$ (pulses commute with system Hamiltonian) and $\sum_{\ell=0}^{L-1} g_\ell H_{SB} g_\ell^\dagger = 0$ (decoupling condition). Let $D$ denote the evolution generated by the applying the DD sequence*

$$D = \prod_{\ell=0}^{L-1} g_\ell \exp[-i(H_0 + H_{SB})\tau] g_\ell^\dagger, \quad g_\ell \in G. \tag{S121}$$

*Then the performance of the probabilistic DD defined by $\mathcal{R}(\rho) = \frac{1}{|G|} \sum_{g \in G} (g D g^\dagger) \rho (g D^\dagger g^\dagger)$ is*

$$\|\mathcal{R} - \mathcal{U}\|_\diamond \leq \|D - U_0\|^2 + J^2 T^2 \left[ 1 + \frac{T}{2}(2\beta + J) \right]. \tag{S122}$$

*Proof.* Lemma S2 states that

$$\|\mathcal{R} - \mathcal{U}\|_\diamond \leq \max_g \|g D g^\dagger - U_0\|^2 + 2 \left\| \frac{1}{|G|} \sum_{g \in G} g D g^\dagger - U_0 \right\|. \tag{S123}$$

Because by assumption $g \in G$ commutes with $H_0$, the first term is simply $\|D - U_0\|^2$. To bound the second term, note that

$$g D g^\dagger = \prod_{l=1}^{L} \exp[-i(H_0 + g g_\ell H_{SB} g_\ell^\dagger g^\dagger)\tau]. \tag{S124}$$

Define $V(t) = g_\ell H_{SB} g_\ell^\dagger$ for $\ell\tau \leq t < (\ell+1)\tau$. Then $g D g^\dagger$ is generated by $H_0 + g V(t) g^\dagger$. Define

$$\frac{d}{dt} U_g = -i[H_0 + g V(t) g^\dagger] U_g, \quad U_g(0) = I, \tag{S125}$$

so that $U_g(T) = g D g^\dagger$. For every $t$, we have

$$U_0^\dagger(t) U_g(t) = I - i \int_0^t U_0^\dagger(t_1) g V(t_1) g^\dagger U_g(t_1) dt_1$$
$$= I - i \int_0^t U_0^\dagger(t_1) g V(t_1) g^\dagger U_0(t_1) dt_1 - \int_0^t dt_1 U_0^\dagger(t_1) g V(t_1) g^\dagger U_0(t_1) \int_0^{t_1} dt_2 U_0^\dagger(t_2) g V(t_2) g^\dagger U_g(t_2), \tag{S126}$$

which gives

$$\sum_{g \in G} [U_0^\dagger(T) U_g(T) - I] = -i \sum_{g \in G} \int_0^T U_0^\dagger(t_1) g V(t_1) g^\dagger U_0(t_1) dt_1$$

$$- \sum_{g \in G} \int_0^T dt_1 U_0^\dagger(t_1) g V(t_1) g^\dagger U_0(t_1) \int_0^{t_1} dt_2 U_0^\dagger(t_2) g V(t_2) g^\dagger U_g(t_2). \tag{S127}$$

The key observation is that the first term on the RHS is zero since for all $t$,

$$\sum_{g \in G} g V(t) g^\dagger = \sum_{g \in G} g g_\ell H_{SB} g_\ell^\dagger g^\dagger = \sum_{g \in G} g H_{SB} g^\dagger = 0. \tag{S128}$$

This already proves the statement of the theorem since the second term has two $V(t)$ operators and therefore it is $\mathcal{O}(J^2)$. To derive a concrete upperbound we note that:

$$\left\| \frac{1}{|G|} \sum_{g \in G} U_g(T) - U_0(T) \right\| = \frac{1}{|G|} \left\| \sum_{g \in G} \int_0^T dt_1 U_0^\dagger(t_1) g V(t_1) g^\dagger U_0(t_1) \int_0^{t_1} dt_2 U_0^\dagger(t_2) g V(t_2) g U_g(t_2) \right\|$$

$$\leq \max_{g \in G} \left\| \int_0^T dt_1 U_0^\dagger(t_1) g V(t_1) g^\dagger U_0(t_1) \int_0^{t_1} dt_2 U_0^\dagger(t_2) g V(t_2) g^\dagger U_g(t_2) \right\|$$

$$(\text{becasue} [g, H_0] = 0) \quad = \max_{g \in G} \left\| \int_0^T dt_1 U_0^\dagger(t_1) V(t_1) U_0(t_1) \int_0^{t_1} dt_2 U_0^\dagger(t_2) V(t_2) g^\dagger U_g(t_2) g \right\|$$

$$= \left\| \int_0^T dt_1 U_0^\dagger(t_1) V(t_1) U_0(t_1) \int_0^{t_1} dt_2 U_0^\dagger(t_2) V(t_2) U_I(t_2) \right\|. \tag{S129}$$

where $U_I$ is the unitary generated by $H_0 + V(t)$ (corresponding to $g$ in $U_g$ being the Identity). The last equation holds because similar to $U_I$ the unitary $g^\dagger U_g g$ is generated by $H_0 + V$ (see Eq. (S125)), and since $g^\dagger U_g(0) g = U_I(0) = 0$ we have $g^\dagger U_g(t) g = U_I(t)$ . Finally, following Eq. (S44) in the proof of Theorem S2 , we obtain

$$\|\mathcal{R} - \mathcal{U}\|_\diamond \leq \|U_0 - D\|^2 + T \cdot \max_{t \in [0,T]} \|V(t) S(t)\| + \frac{T^2}{2} J(2\beta + J) \cdot \max_{t \in [0,T]} \|S(t)\| \tag{S130}$$

$$\leq \|U_0 - D\|^2 + JT \left[ 1 + \frac{T}{2}(2\beta + J) \right] \cdot \max_{t \in [0,T]} \|S(t)\| \tag{S131}$$

$$\leq \|U_0 - D\|^2 + J^2 T^2 \left[ 1 + \frac{T}{2}(2\beta + J) \right]. \tag{S132}$$

The second term in the RHS is $\mathcal{O}(J^2)$, hence the proof is completed. Note that as discussed earlier, in many scenarios the upperbound $\|S(t)\| \leq JT$ can be improved to $\|S(t)\| = \mathcal{O}(J\tau)$.

$\square$

**Remark S1.** *The results can be extended to the setting where pulse intervals $\{\tau_\ell\}$ are not identical, by defining the decoupling condition to be $\sum_{\ell=0}^{L-1} g_\ell H_{SB} g_\ell^\dagger \tau_\ell = 0$. Define $V(t) = g_\ell H_{SB} g_\ell^\dagger$ for $\sum_{\ell' < \ell} \tau_{\ell'} \leq t < \sum_{\ell' \leq \ell} \tau_\ell$. Following the proof of Theorem S5, we then have $gDg^\dagger$ is generated by $H_0 + gV(t)g^\dagger$. For this $V(t)$, one can verify that Eq. (S128) still holds. The rest of the proof is the same.*

**Remark S2.** *The decoupling condition, $\sum_{\ell=0}^{L-1} g_\ell H_{SB} g_\ell^\dagger = 0$, can be relaxed to $\sum_{\ell=0}^{L-1} g_\ell H_{SB} g_\ell^\dagger \propto I_S \otimes B$ where $B$ is an arbitrary bath operator, as the identity component introduces a phase shift to the system's evolution. A suitable metric, such as operator norm optimized for relative phase $(\min_\phi \|U_0 - e^{i\phi} U_1\|)$, can be used to account for the relative phase between $U_0$ and $U_1$.*

## IV. PROOF OF THEOREM 2 AND RELEVANT EXAMPLES

In this section, we provide a proof of Theorem 2, demonstrating that when we focus on the error within the system Hilbert space, the sequence-randomized DD protocol provides a full quadratic improvement over its deterministic counterpart.

We consider the system-bath Hamiltonian without any system Hamiltonian (i.e. $H_S = 0$) $H = I_S \otimes H_B + H_{SB}$. Let an initial state be $\rho = \rho_S(0) \otimes \rho_B \in \mathcal{H}_S \otimes \mathcal{H}_B$. After applying the deterministic DD sequence described in Eq.(11) and its sequence-randomized version described in Eq.(12), and tracing out the environment, the reduced density operators for the system Hilbert space become:

$$\rho_S(T) = \mathrm{tr}_B(D\rho D^\dagger), \quad \rho_S^{\mathrm{ran}}(T) = \mathrm{tr}_B(\mathcal{R}(\rho)), \tag{S133}$$

respectively. The subsystem errors of interest are given by $\frac{1}{2}\|\rho_S(T) - \rho_S(0)\|_1$ and $\frac{1}{2}\|\rho_S^{\mathrm{ran}}(T) - \rho_S(0)\|_1$.

The proof of Eq.(13) is a straightforward generalization of the proof presented in Refs.[9, 10]. For completeness, and to highlight where the quadratic advantage of our scheme arises, we begin the proof of Theorem 2 by giving detailed proof of the bound.

**Theorem S6** (Randomized DD error bound for system Hilbert space; Restatement of Theorem 2). *Consider the expansion of a given deterministic DD protocol $D = \sum_{P \in \mathcal{P}} P \otimes E_P$ where $\mathcal{P}$ is the set of required (multi-qubit) Pauli operators on the system Hilbert space, and assume $\sum_{P \in \mathcal{P} \setminus \{I\}} \|E_P\| \leq 1$. Choose a decoupling group $G$ that satisfies $\forall P \in \mathcal{P} : \sum_{g \in G} gPg^\dagger = 0$ and define the corresponding sequence-randomized DD as $\mathcal{R}(\rho) = \frac{1}{|G|} \sum_{g \in G}(gDg^\dagger)\rho(gD^\dagger g^\dagger)$. Then we have*

$$\forall \rho_S(0) : \quad \frac{1}{2}\|\rho_S(T) - \rho_S(0)\|_1 \leq 2 \sum_{P \in \mathcal{P} \setminus \{I\}} \|E_P\| \tag{S134}$$

$$\forall \rho_S(0) : \quad \frac{1}{2}\|\rho_S^{ran}(T) - \rho_S(0)\|_1 \leq 2\Big(\sum_{P \in \mathcal{P} \setminus \{I\}} \|E_P\|\Big)^2. \tag{S135}$$

*Proof.* Consider the decomposition of the deterministic DD evolution $D$ into Pauli operators supported on the $n$-qubit system Hilbert space $\mathcal{H}_S$:

$$D = \sum_{P \in \mathcal{P}} P \otimes E_P, \tag{S136}$$

where $\mathcal{P} = \{P\}$ is the set of $n$-qubit Pauli operators, and $\{E_P\}$ is the associated operator defined on $\mathcal{H}_B$. The unitarity of $D$, $DD^\dagger = I$, ensures that $\forall P$, $\|E_P\| \leq 1$ and in addition

$$DD^\dagger = \sum_P I_S \otimes E_P E_P^\dagger + \sum_{P \neq Q} PQ \otimes E_P E_Q^\dagger = I_S \otimes I_B, \tag{S137}$$

which implies that

$$\sum_P E_P E_P^\dagger = I_B. \tag{S138}$$

Given any initial state of $\rho = \rho_S \otimes \rho_B$, the time-evolved state of the deterministic DD protocol becomes

$$\begin{aligned}
\rho_S(T) = \mathrm{tr}_B(D\rho_S \otimes \rho_B D^\dagger) &= \sum_{P,Q} \mathrm{tr}(E_Q^\dagger E_P \rho_B) P \rho_S Q \\
&= \sum_P \mathrm{tr}(E_P^\dagger E_P \rho_B) P \rho_S P + \sum_{P \neq Q} \mathrm{tr}(E_Q^\dagger E_P \rho_B) P \rho_S Q \\
&= \mathrm{tr}(E_I^\dagger E_I \rho_B) \rho_S + \sum_{P \in \mathcal{P} \setminus \{I\}} \mathrm{tr}(E_P^\dagger E_P \rho_B) P \rho_S P + \sum_{P \neq Q} \mathrm{tr}(E_Q^\dagger E_P \rho_B) P \rho_S Q.
\end{aligned} \tag{S139}$$

Now we proceed to bound:

$$\frac{1}{2}\|\rho_S(T) - \rho_S(0)\|_1 = \frac{1}{2}\left\| (\mathrm{tr}(E_I^\dagger E_I \rho_B) - 1)\rho_S + \sum_{P \in \mathcal{P} \setminus \{I\}} \mathrm{tr}(E_P^\dagger E_P \rho_B) P \rho_S P + \sum_{P \neq Q} \mathrm{tr}(E_Q^\dagger E_P \rho_B) P \rho_S Q \right\|_1. \tag{S140}$$

Using the first equation of Eq.(S138) we get $\text{tr}(E_I^\dagger E_I \rho_B) - 1 = -\sum_{P \in \mathcal{P}\setminus\{I\}} \text{tr}(E_P^\dagger E_P \rho_B)$. Plugging this into the equation above and using Hölder's inequality give

$$
\begin{aligned}
\frac{1}{2} \|\rho_S(T) - \rho_S(0)\|_1 &\leq \sum_{P \in \mathcal{P}\setminus\{I\}} \|E_P\|^2 + \frac{1}{2} \sum_{P \neq Q} \|E_P\| \cdot \|E_Q\| \\
&= \sum_{P \in \mathcal{P}\setminus\{I\}} \|E_P\|^2 + \frac{1}{2} \sum_{\substack{P \neq Q, \\ P,Q \in \mathcal{P}\setminus\{I\}}} \|E_P\| \cdot \|E_Q\| + \sum_{P \in \mathcal{P}\setminus\{I\}} \|E_P\| \cdot \|E_I\| \\
&\leq \left( \sum_{P \in \mathcal{P}\setminus\{I\}} \|E_P\| \right)^2 + \sum_{P \in \mathcal{P}\setminus\{I\}} \|E_P\| \\
&\leq 2 \sum_{P \in \mathcal{P}\setminus\{I\}} \|E_P\|,
\end{aligned}
\tag{S141}
$$

where from the second to the third line, we have used $\|E_I\| \leq 1$, and in the last inequality we used the assumption that $\sum_{P \in \mathcal{P}\setminus\{I\}} \|E_P\| \leq 1$. This proves the bound in Eq.(13).

Now, let us consider the state obtained from the randomized DD protocol:

$$
\rho_S^{\text{ran}}(T) = \text{tr}_B(\mathcal{R}(\rho)), \quad \mathcal{R}(\rho) = \frac{1}{|G|} \sum_{g \in G} (gDg^\dagger)\rho(gDg^\dagger).
\tag{S142}
$$

For an initial state $\rho = \rho_S \otimes \rho_B$, this state can be expressed as

$$
\rho_S^{\text{ran}}(T) = \frac{1}{|G|} \sum_{g \in G} \text{tr}_B(gDg^\dagger \rho_S \otimes \rho_B g D^\dagger g^\dagger)
\tag{S143}
$$

$$
= \frac{1}{|G|} \sum_{g \in G} \sum_{P \in \mathcal{P}} \text{tr}(E_P^\dagger E_P \rho_B) gPg^\dagger \rho_S gPg^\dagger + \frac{1}{|G|} \sum_{g \in G} \sum_{P \neq Q} \text{tr}(E_Q^\dagger E_P \rho_B) gPg^\dagger \rho_S gQg^\dagger.
\tag{S144}
$$

The assumption $\forall P \in \mathcal{P} : \sum_{g \in G} gPg^\dagger = 0$ guarantees that, in the second term of the RHS of Eq.(S144), whenever either $P = I$ or $Q = I$, we have (without loss of generality we choose $Q = I$)

$$
\sum_{g \in G} \sum_{P \in \mathcal{P}\setminus\{I\}} \text{tr}(E_I^\dagger E_P \rho_B) gPg^\dagger \rho_S = 0.
\tag{S145}
$$

Therefore the state obtained from the randomized DD protocol can be expressed as follows:

$$
\rho_S^{\text{ran}}(T) = \sum_{P \in \mathcal{P}\setminus\{I\}} \text{tr}(E_P^\dagger E_P \rho_B) \cdot \frac{1}{|G|} \sum_{g \in G} gPg^\dagger \rho_S gPg^\dagger + \sum_{\substack{P \neq Q \\ P,Q \in \mathcal{P}\setminus\{I\}}} \text{tr}(E_Q^\dagger E_P \rho_B) \cdot \frac{1}{|G|} \sum_{g \in G} gPg^\dagger \rho_S gQg^\dagger.
\tag{S146}
$$

Finally, using $\text{tr}(E_I^\dagger E_I \rho_B) = 1 - \sum_{P \in \mathcal{P}\setminus\{I\}} \text{tr}(E_P^\dagger E_P \rho_B)$ and Hölder's inequality again, we obtain

$$
\frac{1}{2} \|\rho_S^{\text{ran}}(T) - \rho_S(0)\|_1 \leq 2 \sum_{P \in \mathcal{P}\setminus\{I\}} \|E_P\|^2 + \sum_{\substack{P \neq Q \\ P,Q \in \mathcal{P}\setminus\{I\}}} \|E_P\| \cdot \|E_Q\| \leq 2 \left( \sum_{P \in \mathcal{P}\setminus\{I\}} \|E_P\| \right)^2.
\tag{S147}
$$

$\square$

Below, we consider two applications of this theorem.

## A. Example 1: Uhrig DD (UDD); Proof of Corollary 1

The goal of UDD [11] is to suppress pure dephasing noise, which can be characterized by the following system-bath Hamiltonian:

$$
H = H_B + H_{SB} = I_S \otimes B_I + Z \otimes B_Z.
\tag{S148}
$$

In the $K$-th order UDD, $K$ (or $K + 1$ depending on the parity of $K$) number of $X$ pulses are applied at specific time $t_j = T\delta_j$ where $T$ is the total time of the DD sequence and (cf. Eq.(4) in Ref.[12])

$$\delta_j = \begin{cases} \sin^2\left(\frac{j\pi}{2K+2}\right), & j = 1, 2, \ldots, K. \quad (\text{even } K) \\ \sin^2\left(\frac{j\pi}{2K+2}\right), & j = 1, 2, \ldots, K + 1. \quad (\text{odd } K) \end{cases} \tag{S149}$$

Since $X$ pulses only flip the sign of $H_{SB}$ term and $Z^2 = I$, it is clear that only the odd powers of $B_Z$ contribute to the dephasing, while the even powers do not. Hence, the UDD protocol can be expressed as:

$$D = I \otimes E_I + Z \otimes E_Z, \tag{S150}$$

where $E_Z$ contains all terms in the form of $(Z \otimes B_Z)^k$ for odd $k$ and $E_I$ contains the rest of the remaining term. With the aforementioned carefully designed sequence of $K$ pulses, it has been shown that dephasing error can be reduced to $\mathcal{O}(T^{K+1})$. More precisely, this refers to the following condition [12]

$$\|E_Z\| = \mathcal{O}(T^{K+1}). \tag{S151}$$

From Eq.(S134) we have (note that the following result for UDD was also demonstrated in Ref.[9])

$$\frac{1}{2}\|\rho_S(T) - \rho_S(0)\| = \mathcal{O}(T^{K+1}). \tag{S152}$$

Thus, UDD is considered optimal, as adding an extra pulse decreases the error by one order of magnitude.

To construct the randomized sequences we note that $\mathcal{P} = \{I, Z\}$, and we choose the decoupling group $G = \{I, X\}$ which satisfies $\forall P \in \mathcal{P}: \sum_{g \in G} gPg^\dagger = 0$. The state obtained from our sequence-randomized protocol becomes

$$\rho_S^{\text{ran}}(T) = \text{tr}_B\left(\frac{1}{2}D\rho D^\dagger + \frac{1}{2}XDX\rho XD^\dagger X\right), \tag{S153}$$

and Theorem S6 guarantees

$$\frac{1}{2}\|\rho_S^{\text{ran}}(T) - \rho_S(0)\| = \mathcal{O}(T^{2K+2}). \tag{S154}$$

which completes the proof of Corollary 1.

**Explicit description of pulses for randomized UDD:** The $K$-th order randomized UDD protocol implements $XDX$ and $D$ with equal probability. Here we explicitly write down the list of pulses.

In $K$-th order deterministic UDD, $X$ pulses are applied at specific time $t_j = T\delta_j$ according to Eq.(S149). The modified sequence $XDX$ apply $K + 2$ (or $K + 1$ depending on the parity of $K$) at $t_j = T\delta'_j$ where

$$\delta'_j = \begin{cases} \sin^2\left(\frac{j\pi}{2K+2}\right), & j = 0, 1, 2, \ldots, K, K + 1. \quad (\text{even } K) \\ \sin^2\left(\frac{j\pi}{2K+2}\right), & j = 0, 1, 2, \ldots, K. \quad (\text{odd } K) \end{cases} \tag{S155}$$

For any given order $K$, our sequence-randomized UDD protocol randomly selects between using the given original sequence $D$ (which is generated by the pulse sequence described in Eq.(S149)) and its corresponding modified sequence (which is generated by the pulse sequence described in Eq.(S155)).

Therefore, for even $K$, at most two additional $X$ pulses are required, while for odd $K$, the total number of pulses remains the same. Let the total number of pulses $L$ be the same for both deterministic and randomized UDD protocols. In this case, the deterministic UDD achieves an error scaling of $\mathcal{O}(T^{L+1})$, while the randomized UDD achieves $\mathcal{O}(T^{2L-2})$, which implies that the randomized UDD outperforms its deterministic counterpart whenever $L > 3$.

### B. Example 2: Quadratic DD (QDD)

QDD [13] extends UDD to protect against general single-qubit decoherence. The key idea of QDD is to nest two UDD sequences with different pulse types (e.g., both $X$ and $Z$ pulses) and different number of pulses. We can express the deterministic QDD protocol as

$$D = I \otimes E_I + X \otimes E_X + Y \otimes E_Y + Z \otimes E_Z, \tag{S156}$$

where it is proven that [14, 15]

$$\|E_\sigma\| = \mathcal{O}(T^{K+1}), \quad \sigma \in \{X, Y, Z\}. \tag{S157}$$

Then Eq.(S134) gives

$$\frac{1}{2}\|\rho_S(T) - \rho_S(0)\|_1 = \mathcal{O}(T^{K+1}). \tag{S158}$$

To construct the randomized QDD, note that $\mathcal{P} = \{I, X, Y, Z\}$ and therefore we choose the decoupling group $G = \{I, X, Y, Z\}$. The time-evolved state under the randomized DD protocol becomes

$$\rho_S^{\mathrm{ran}}(T) = \mathrm{tr}_B\left(\frac{1}{4}D\rho D^\dagger + \frac{1}{4}D_X\rho D_X^\dagger + \frac{1}{4}D_Y\rho D_Y^\dagger + \frac{1}{4}D_Z\rho D_Z^\dagger\right), \quad D_\sigma \equiv \sigma D \sigma^\dagger. \tag{S159}$$

Using Theorem S6 we have

$$\frac{1}{2}\|\rho_S^{\mathrm{ran}}(T) - \rho_S(0)\|_1 = \mathcal{O}(T^{2K+2}). \tag{S160}$$

## V. ADDITIONAL NUMERICAL SIMULATIONS AND BACKGROUND FOR EXPERIMENTAL IMPLEMENTATION

In this section we discuss the experimental implementation through examples, provide more numerical simulations, and also provide examples of using the scheme to suppress noise on qudit systems.

### A. Background on the choice of error measures and the implementation of $\mathcal{R}$

We first provide the background to connect the mathematical derivations in the paper to experimental realizations. In particular we review the importance of using trace and diamond distance in measuring errors, and the fact that implementation of the randomized DD in compilation occur no extra sampling cost to estimate observables.

**Remark 1 (on trace and diamond distances):** In proving the theorems we used the trace and diamond norm as the measure of error. These choices are due to their precise operational interpretations in distinguishing quantum states and quantum channels, respectively [16, 17]. This is in contrast to other measures such as the infidelity of mixed states that can be loose in bounding the error in estimating general observables [16].

For any two density matrices $\rho$ and $\sigma$, and a Hermitian operator $O$, Hölder's inequality states:

$$|\mathrm{tr}(O\rho) - \mathrm{tr}(O\sigma)| \leq 2\|O\|\left(\frac{1}{2}\|\rho - \sigma\|_1\right), \tag{S161}$$

and therefore a bound on the trace distance guarantees the error for any observable. Importantly this upperbound is saturated when the measurement is chosen to be the Helstrom measurement [16]. In this sense, the trace distance is the fundamental measure of distinguishability of quantum states. Hence, Theorem 2 in the main text demonstrates that our randomized protocol can estimate the expectation value of any observable $O$ with better guarantee compared to the deterministic counterpart.

Similarly, the diamond distance is the fundamental measure of distinguishability of quantum channels [17]. Therefore, the bound on the diamond norm from Theorem 1 also bounds the trace distance for any input state

$$\forall \rho, \mathcal{E}_1, \mathcal{E}_2 : \quad \|\mathcal{E}_1(\rho) - \mathcal{E}_2(\rho)\|_1 \leq \|\mathcal{E}_1 - \mathcal{E}_2\|_\diamond, \tag{S162}$$

which provides the guarantee on the performance of the randomized protocol in estimating any observable.

In contrast, a measure such as infidelity can be a (quadratically) loose measure in distinguishing mixed states (and hence bounding the errors in the estimating observables). For example, if the infidelity of a pure state $|\psi\rangle$ and a mixed state $\rho$ is $\epsilon$, i.e. $1 - \langle\psi|\rho|\psi\rangle = \epsilon$, then the only bound on distinguishibility is $\epsilon \leq \frac{1}{2}\|\rho - |\psi\rangle\langle\psi|\|_1 \leq \sqrt{\epsilon}$ [16]. The bounds are tight. To see this consider $|\psi\rangle\langle\psi| = |0\rangle\langle0|$, $\rho_1 = |\phi\rangle\langle\phi|$ where $|\phi\rangle = \sqrt{1-\epsilon}|0\rangle + \sqrt{\epsilon}|1\rangle$, and $\rho_2 = (1-\epsilon)|0\rangle\langle0| + \epsilon|1\rangle\langle1|$. Both $\rho_1$ and $\rho_2$ have $\epsilon$ infidelity with $|0\rangle\langle0|$. But $\frac{1}{2}\||\psi\rangle\langle\psi| - \rho_1\|_1 = \sqrt{\epsilon}$ and $\frac{1}{2}\||\psi\rangle\langle\psi| - \rho_2\|_1 = \epsilon$. Hence the infidelity of mixed states can be quadratically loose in bounding the error in estimating general observables. This is particularly important in our setup since the randomized protocol provides a quadratic (in the coupling to environment) improvement in reducing the error of estimating observable.

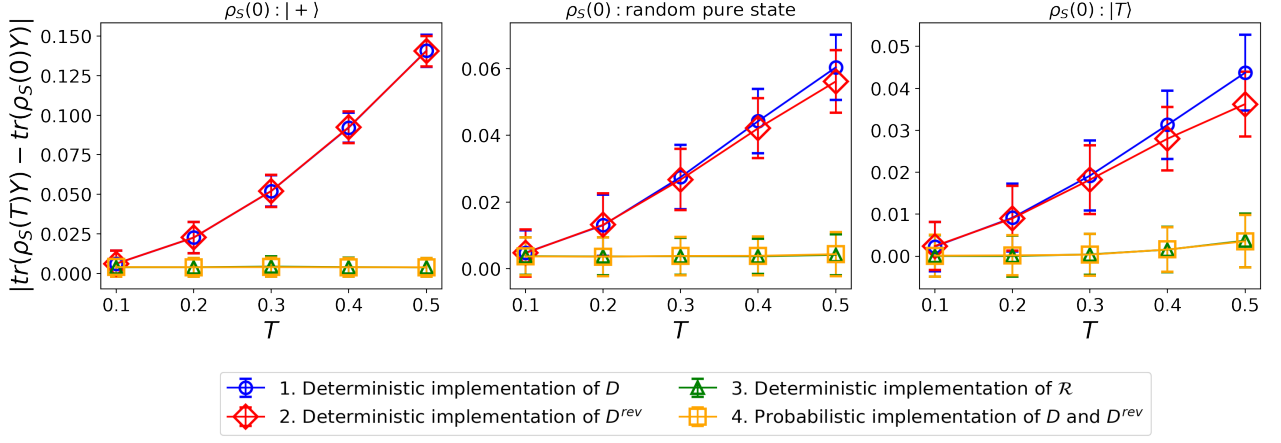

FIG. S1: Performance on estimating an expectation values of $Y$ observable for the four different scenarios on choosing the DD protocol. For each evolution time $T$, we fixed the number of measurements (i.e. shots) to be $M = 40000$ and repeated the experiment $N = 1000$ times to report the mean and standard deviation $\sigma$, with the $2\sigma$ plotted as the error bar. We consider three different initial states: $\rho = \rho_S(0) \otimes \rho_B$, where $\rho_S(0)$ is selected as $|+\rangle\langle+|$ (leftmost panel), a random pure state with (middle panel), and $|T\rangle\langle T|$ with $|T\rangle = \cos(\beta)|0\rangle + e^{i\pi/4}\sin(\beta)|1\rangle$ (where $\beta = \arccos(1/\sqrt{3})/2$) being the $T$-type magic state [18] (rightmost panel). Here $\rho_B$ is chosen as a random pure state for all cases.

**Remark 2: (Implementation of $\mathcal{R}$ with no sampling overhead)** The randomized DD protocol $\mathcal{R}$, which is a probabilistic mixture of unitary operation (with equal weights), can be physically implemented by randomly choosing one of the deterministic DD protocols for each compilation of the quantum circuit and averaging the results over multiple runs. For estimating expectation values of any observable, this random sampling process does not introduce any additional sampling costs, allowing expectation values to be estimated as if $\mathcal{R}$ were implemented. First note that there is no cost associated with generating the classical randomness since the probabilities in the mixture are all uniform and hence easy to generate. Also all the statistics of these two scenarios to estimate an observable is the same. (This is a manifestation of the fact that there is no physical process that can distinguish $\rho = \sum_i p_i \rho_i$ from the probabilistic implementations of $p_i$.)

To see this, let $\rho$ be an arbitrary initial state and $O$ an observable we want to measure. Consider the spectral decomposition $O = \sum_{i=1}^{F} a_i \Pi_i$. In the case where we assume a direct implementation of $\mathcal{R}$ (without sampling), the outcomes are independent and identically distributed random variables $X : \{1, \ldots, F\} \to \mathbb{R}$ with the associated probability of $\Pr(X = a_i) = \text{tr}(\Pi_i \mathcal{R}(\rho)) = p_i$. Now, consider a different scenario where, in each experiment, we randomly select and implement a unitary $g_j D g_j^\dagger$ (with $g_j \in G$) uniformly at random. In this case, the random variable $Y : \{1, \ldots, F\} \times \{1, \ldots, |G|\} \to \mathbb{R}$ has an associated probability of $\Pr(Y = a_i) = \text{tr}(\Pi_i \rho_j) \cdot \frac{1}{|G|} = q_{i,j}$, where $\rho_j = (g_j D g_j^\dagger) \rho (g_j D g_j^\dagger)^\dagger$. Noting that $\sum_{j=1}^{|G|} q_{i,j} = p_i$, and that the outcomes of $Y$ are independent of $j$, we conclude that for any function $f(\cdot)$ we have $\mathbb{E}[f(Y)] = \sum_{i,j} f(a_i) q_{i,j} = \sum_i f(a_i) p_i = \mathbb{E}[f(X)]$, for example the two random variables have the same characteristic functions (and mean and variance $\mathbb{E}[Y] = \mathbb{E}[X]$ and $\mathbb{E}[Y^2] = \mathbb{E}[X^2]$). Therefore it is not possible to distinguish the two scenarios using statistics of any such observable $O$.

### B. Steps of the protocol to measure an observable in an experiment

To describe the steps of the protocol in an experiment, consider a single-qubit system interacting with a single-qubit bath:

$$H = I \otimes B_I + J(Z \otimes B_Z),$$ (S163)

where both $B_I$ and $B_Z$ are arbitrary single-qubit operators. Our goal is to measure an expectation value of a given observable $O$, where we account for a finite number of measurement or shots $M$. Similar to the illustrative example presented in the main text, one can consider two deterministic DD protocols defined as

$$D = e^{-iH\tau} X e^{-iH\tau} X, \quad D^{\text{rev}} = X e^{-iH\tau} X e^{-iH\tau}.$$ (S164)


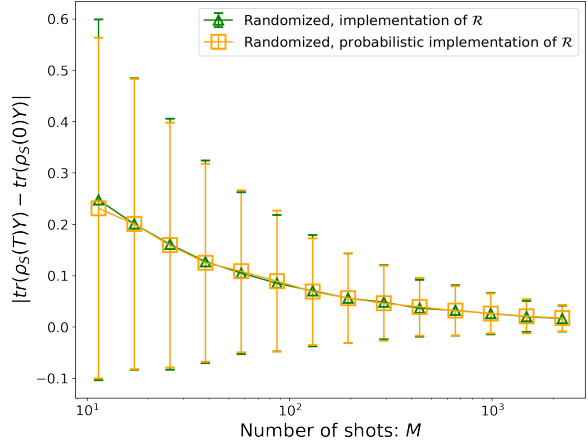

FIG. S2: Performance comparison between implementing the randomized DD protocol $\mathcal{R}$ and implementing the randomized DD protocol by probabilistically applying $D$ and $D^{\text{rev}}$ with equal probability (scenario 3 vs scenario 4) for estimating the expectation value of the $Y$ observable. Here we have fixed parameters $J = 1, T = 0.4$ and considered $\rho_S(0) = |+\rangle \langle +|$. For each fixed number of shots $M$, we repeat the experiment for $N = 1000$ times and report the mean along with the $2\sigma$ errors, where we observe that both the mean and the standard deviation $\sigma$ for both cases are nearly identical.

Then, our sequence-randomized DD protocol is

$$\mathcal{R}(\rho) = \frac{1}{2} D \rho D^{\dagger} + \frac{1}{2} D^{\text{rev}} \rho D^{\text{rev} \dagger}. \tag{S165}$$

To implement the randomized protocol, for each of the $M$ measurements (i.e., shots) we flip a coin and either implement $D$ or $D^{\text{rev}}$ based on the result of the coin. We collect the result of measuring observable $O$ for each shot, and report the average. Then, we repeat this process $N$ times and report both the average and the standard deviation.

Below we present the numerical simulation of this experiment, where we report the result for four scenarios:

1. Every time we run the DD protocol to collect a measurement outcome, we deterministically choose $D$.

2. Every time we run the DD protocol to collect a measurement outcome, we deterministically choose $D^{\text{rev}}$.

3. Every time we run the DD protocol to collect a measurement outcome, we deterministically implement $\mathcal{R}$.

4. Every time we run the DD protocol to collect a measurement outcome, we randomly choose either $D$ or $D^{\text{rev}}$ with equal probability.

The scenario 4 above describes the implementation of our randomized protocol in real experimental setups.

In the numerical simulations, we fixed $J = 1$ and chose both $B_I$ and $B_Z$ as $\alpha X + \beta Y + \gamma Z + \delta I$ with where coefficients are randomly chosen from $[0, 1]$. We attempt to estimate the expectation values of the $O = Y$ observable for different evolution times $T$. Specifically, we estimate the expectation value of $Y$, $\text{tr}(\rho_S(T)Y)$ based on the $M$ binary samples (based on the obtained measurement outcome), where $\rho_S(t)$ is the reduced density operator in the system Hilbert space at time $t$. The number of measurements (shots) is fixed to be $M = 40000$, and all the parameters $B_I, B_Z, \rho_B$ are kept fixed for all the shots. Then, we compute the absolute value difference between this estimated value and the ideal expectation value, $|\text{tr}(\rho_S(T)Y) - \text{tr}(\rho_S(0)Y)|$. We then repeat this procedure $N = 1000$ times to collect a set of the differences, obtain the average and the standard deviation $\sigma$, and plot the mean and $2\sigma$ errors in all of our graphs.

The numerical results are shown in Fig. S1. We consider three different initial states: $\rho = \rho_S(0) \otimes \rho_B$, where $\rho_S(0)$ is chosen as $|+\rangle \langle +|$ (leftmost panel), a random pure state with $\rho_B$ (middle panel), and $|T\rangle \langle T|$ with $|T\rangle = \cos(\beta) |0\rangle + e^{i\pi/4} \sin(\beta) |1\rangle$ (where $\beta = \arccos(1/\sqrt{3})/2$) represents the $T$-type magic state [18] (rightmost panel). We choose $\rho_B$ to be a random pure state for all cases.

As expected, Fig. S1 demonstrates that our randomized protocol estimates the expectation value more accurately compared to the deterministic counterpart, aligning with what our theorems suggest, as discussed in Remark 1 in Section V A. Note that since the variance scales as $1/\sqrt{M} = 1/200$, we expect $2\sigma = 0.01$, which is consistent with the graphs. Also note that scenario 3 and scenario 4 have the same statistical properties when measuring observables, as described in Remark 2 in the same section (see also Fig. S2).

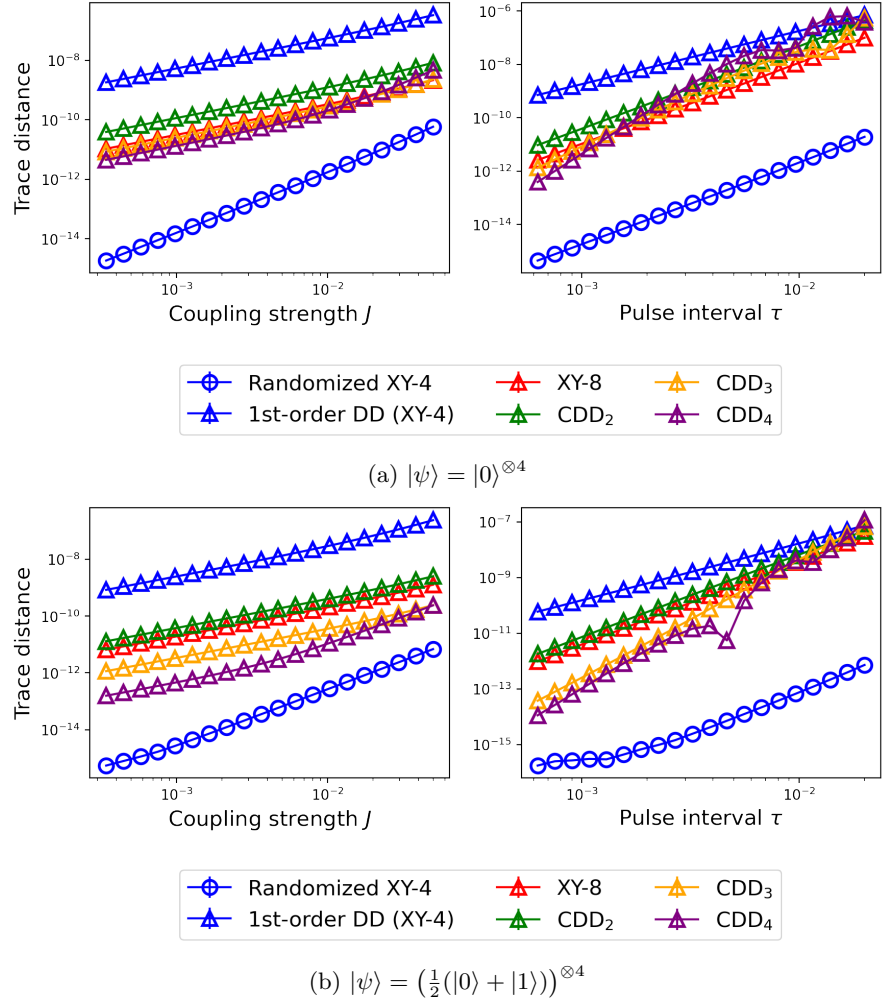

FIG. S3: Performance of deterministic DD protocols ($\text{CDD}_K$ for $K = 1, 2, 3, 4$ (where $\text{CDD}_1 = $ XY4) and XY8) and the sequence-randomized XY4, for a single fixed initial state $\rho = \rho_S \otimes \rho_B$ where $\rho_S = |\psi\rangle \langle\psi|$, and $\rho_B$ is a random pure state.

## C. Numerical simulations with single initial state

As described in Eq.(S55), the XY4 sequence for a single-qubit system can be represented as $f_\tau Y f_\tau X f_\tau Y f_\tau X$, where $f_\tau = e^{-i\tau H}$ is the free evolution of the total Hamiltonian. Note that swapping either $X$ or $Y$ with the $Z$ operation gives the similar performance. Therefore, for an $n$-qubit system interacting with an environment that introduces 1-local noise on every qubit, the corresponding XY4 sequence can be written as $D = f_\tau Y^{\otimes n} f_\tau X^{\otimes n} f_\tau Y^{\otimes n} f_\tau X^{\otimes n}$. The decoupling group in this scenario is $G = \{I^{\otimes n}, X^{\otimes n}, Y^{\otimes n}, Z^{\otimes n}\}$. Given that our randomized DD protocol applies one of the group elements to the original sequence, it selects one of the four pulse sequences at random:

$$D = f_\tau Y^{\otimes n} f_\tau X^{\otimes n} f_\tau Y^{\otimes n} f_\tau X^{\otimes n}, \qquad \text{(XY4 sequence)} \tag{S166}$$

$$D_{X^{\otimes n}} = X^{\otimes n} D_1 X^{\otimes n}, \tag{S167}$$

$$D_{Y^{\otimes n}} = Y^{\otimes n} D_1 Y^{\otimes n}, \tag{S168}$$

$$D_{Z^{\otimes n}} = Z^{\otimes n} D_1 Z^{\otimes n}. \tag{S169}$$

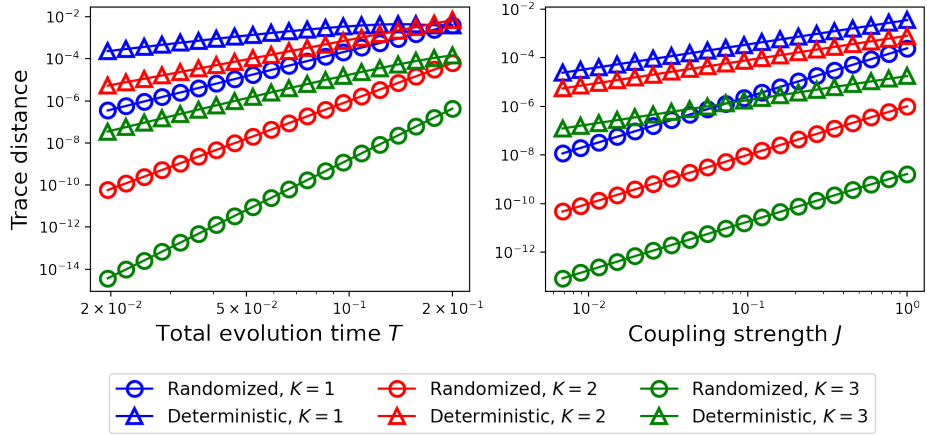

FIG. S4: Performance of deterministic UDD for $K = 1, 2, 3$ order and their corresponding sequence-randomized versions, for a single fixed initial state of $\rho = \rho_S \otimes \rho_B$ where $\rho_S = |\psi\rangle \langle\psi|$ with $|\psi\rangle = \frac{1}{\sqrt{2}}(|0\rangle + |1\rangle)$, and $\rho_B$ is a random pure state.

Some of the pulse operations cancel and merge together, leading to:

$$D = f_\tau Y^{\otimes n} f_\tau X^{\otimes n} f_\tau Y^{\otimes n} f_\tau X^{\otimes n}, \qquad \text{(XY4 sequence)} \tag{S170}$$

$$D_{X^{\otimes n}} = X^{\otimes n} f_\tau Y^{\otimes n} f_\tau X^{\otimes n} f_\tau Y^{\otimes n} f_\tau, \tag{S171}$$

$$D_{Y^{\otimes n}} = Y^{\otimes n} f_\tau Y^{\otimes n} f_\tau X^{\otimes n} f_\tau Y^{\otimes n} f_\tau Z^{\otimes n}, \tag{S172}$$

$$D_{Z^{\otimes n}} = Z^{\otimes n} f_\tau Y^{\otimes n} f_\tau X^{\otimes n} f_\tau Y^{\otimes n} f_\tau Y^{\otimes n}. \tag{S173}$$

Note that $D$ represents the original XY4 sequence, and shifting the first $X^{\otimes n}$ pulse to the end of the sequence results in the $D_{X^{\otimes 4}}$ operation. For $D_{Y^{\otimes 4}}$ (and $D_{Z^{\otimes 4}}$), we modify the initial pulse of the XY4 sequence from $X^{\otimes n}$ to $Z^{\otimes n}$ (and $Y^{\otimes n}$), and introduce an additional pulse operation of $Y^{\otimes n}$ (and $Z^{\otimes n}$) at the end of the sequence. Our randomized XY4 sequence selects and applies one of these pulse sequences with equal probability of $1/4$.

In the main text, we present numerical results comparing the performance of randomized XY4 and other deterministic protocols (e.g., XY4, XY8, and CDD) in Fig.1, along with randomized Uhrig DD and its deterministic counterparts in Fig.2, where the performances are averaged over 20 randomly chosen initial (pure) states. Figs. S3 and S4 demonstrate that the behaviors remain similar when considering a specific, relevant initial state, while keeping the rest of the setup unchanged.

### D. Randomized DD protocols for qudit systems

Our construction of the randomized DD protocol described in Eqs. (7) and (8) is based on a decoupling group and hence is very general. As an example, in this section we show how this general scheme can be directly applied to qudit systems.

The Pauli group is generated by products and tensor products of $X$ and $Z$ operators. The qudit Pauli group, or Heisenberg-Weyl group, is generated by products and tensor products of [19]

$$X_d = \sum_{j=0}^{d-1} |(j+1) \bmod d\rangle \langle j|, \quad Z_d = \sum_{j=0}^{d-1} \omega_d^j |j\rangle \langle j|, \tag{S174}$$

where $\omega_d = e^{2\pi i/d}$ is the $d$th root of the unity. When $d = 2$, it becomes the Pauli group for qubits.

As a concrete example, we focus on single-qutrit system, where the group elements are generated by products of:

$$X_3 = \begin{pmatrix} 0 & 0 & 1 \\ 1 & 0 & 0 \\ 0 & 1 & 0 \end{pmatrix}, \quad Z_3 = \begin{pmatrix} 1 & 0 & 0 \\ 0 & e^{2\pi i/3} & 0 \\ 0 & 0 & e^{4\pi i/3} \end{pmatrix}, \tag{S175}$$

The dephasing noise can be modeled as follows:

$$H = H_{SB} + H_B, \quad H_{SB} = (Z_3 + Z_3^\dagger) \otimes J B_{Z_3}, \quad H_B = I \otimes B_I, \tag{S176}$$

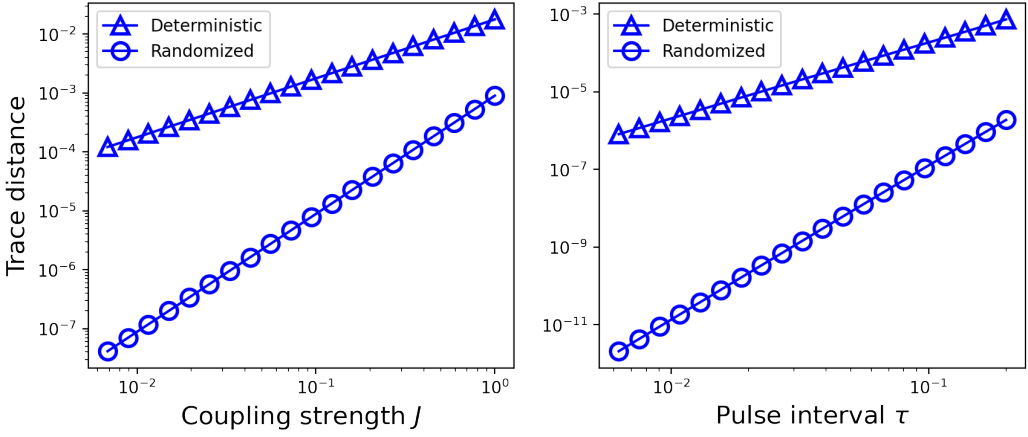

FIG. S5: Performance of deterministic first-order DD and its corresponding sequence-randomized version to suppress a single-axis noise in a single-qutrit system described in Eq.(S176).

where $Z_3^\dagger = Z_3^2$ and $B_{Z_3}, B_I$ are arbitrary $3 \times 3$ operators. The decoupling group for this case is given by $G = \{I, X_3, X_3^2\}$ [20]. Therefore, aligning with our notation in Eq.(6), the deterministic DD is given by

$$D = e^{-iH\tau}(X_3 e^{-iH\tau} X_3^\dagger)((X_3^2) e^{-iH\tau} (X_3^2)^\dagger) \quad (= e^{-iH\tau} X_3 e^{-iH\tau} X_3 e^{-iH\tau} X_3), \tag{S177}$$

which is equivalent to applying 3 $X_3$ pulses with equal pulse interval. Note that to obtain the last equality, we used: $(X_3^2)^\dagger = X_3 = X_3^2 X_3^\dagger$.

Then, our sequence-randomized DD protocol then becomes

$$\mathcal{R}(\rho) = \frac{1}{3} D\rho D^\dagger + \frac{1}{3} D_{X_3} \rho D_{X_3}^\dagger + \frac{1}{3} D_{X_3^2} \rho D_{X_3^2}^\dagger, \quad D_\sigma := \sigma D \sigma^\dagger. \tag{S178}$$

In Fig.S5, for a fixed initial state $\rho = \rho_S \otimes \rho_B$ (where both $\rho_S$ and $\rho_B$ chosen as random pure states), we again plot the trace distance (in the system Hilbert space) between the ideal state and the states evolved over time under the deterministic and randomized DD protocols. As expected, the randomized DD demonstrates an advantage similar to the qubit case.

### E. De-randomization

The simple form of the randomized sequences makes their analysis straightforward. Their implementation in experiments also incurs no extra cost, given that the sequences are chosen uniformly at random, allowing for easy generation of classical randomness. Interestingly, the substantially improved performance of the randomized protocols hints at the existence of new deterministic protocols with better performance compared to the existing ones. The key insight provided by this randomized scheme is the advantage of implementing and averaging several carefully designed (deterministic) DD sequences, rather than using the same (deterministic) DD sequence repeatedly.

For example, consider the example presented in the main text with

$$D = e^{-iH\tau} X e^{-iH\tau} X, \quad D^{\text{rev}} = X e^{-iH\tau} X e^{-iH\tau}. \tag{S179}$$

where $H = H_0 + H_{SB}$, and the sequence-randomized DD protocol defined as

$$\mathcal{R}(\rho) = \frac{1}{2} D\rho D^\dagger + \frac{1}{2} D^{\text{rev}} \rho D^{\text{rev}\dagger} \tag{S180}$$

The goal is to use $M$ measurements (i.e., shots) to estimate an observable.

One example of a deterministic protocol approximating Eq. (S180) is to choose the most likely sequence: to deterministically apply the pulse sequence $D$ for half of the shots ($\frac{M}{2}$ shots) and $D^{\text{rev}}$ for the remaining half. The expected value is reported as the average of the $M$ shots.

To reiterate the reason for a better performance of averaging between applications of $D$ and $D^{\text{rev}}$ from the perspective of estimating observables, consider estimating the expected value of an observable $O$ and assume $H = H_0 + JH_{SB}$.

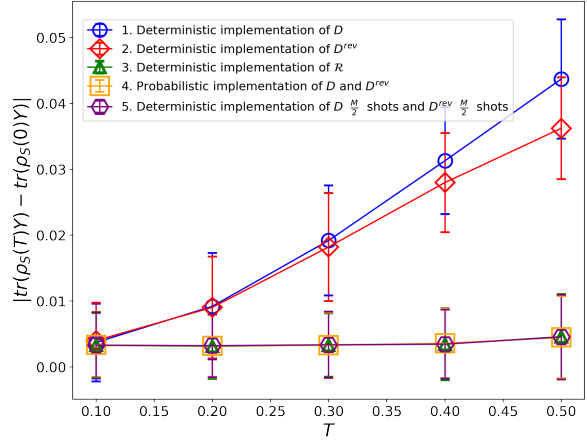

FIG. S6: Performance of the new (de-randomized) deterministic approach, where $D$ is applied for exactly $M/2$ shots and $D^{\text{rev}}$ for the remaining $M/2$ shots. The setup is the same as that used for the rightmost panel of Fig.S1 (measuring the expectation value of $Y$ observable, initial state being $\rho_S(0) = |T\rangle \langle T|$, $M = 40000$ shots per experiments, $N = 1000$ repeated experiments, and etc). Note that the data points for the first four scenarios are identical to those in the rightmost panel of Fig.S1. As expected, the new deterministic protocol, shown by purple data points, closely approximates the randomized DD protocols and thus outperforms the two deterministic protocols.

An expansion of the state $D\rho(0)D^\dagger$ in $J$ yields $e^{-iH_0\tau}\rho(0)e^{+iH_0\tau} + JE_1 + \mathcal{O}(J^2)$, where the exact form of $E_1$ is not important but easy to calculate (see Section I of SM). Hence

$$\text{tr}(OD\rho(0)D^\dagger) - \text{tr}(O\rho_{\text{ideal}}) = JE_1 + \mathcal{O}(J^2), \tag{S181}$$

where $\rho_{\text{ideal}}$ refers to the ideal quantum state time-evolved under $H$ with $H_{SB} = 0$.

The expansion in for $D^{\text{rev}}$ is similar but with changing $J$ to $-J$:

$$\text{tr}(OD^{\text{rev}}\rho(0)(D^{\text{rev}})^\dagger) - \text{tr}(O\rho_{\text{ideal}}) = -JE_1 + \mathcal{O}(J^2), \tag{S182}$$

and therefore averaging the two estimators cancels out the $\mathcal{O}(J)$ bias term and the remaining error is $\mathcal{O}(J^2)$.

Also note that the bound on the trace distance, also bounds the variances as well. In particular, it is straightforward to show that $|\text{Var}(O)_{\mathcal{R}(\rho)} - \text{Var}(O)_{\rho_{\text{ideal}}}| = \mathcal{O}(J^2)$ but $|\text{Var}(O)_{D\rho(0)D^\dagger} - \text{Var}(O)_{\rho_{\text{ideal}}}| = \mathcal{O}(J)$.

Under the same setup as in the rightmost panel of Fig.S1 (i.e., initial state of $\rho_S(0) = |T\rangle \langle T|$, $M = 40000$ shots per experiment, and $N = 1000$ repeated experiments, etc), we deterministically applied $D$ for $M/2 = 20000$ shots and $D^{\text{rev}}$ for the remaining 20000 shots. As shown in the purple data points in Fig. S6, we observe that the performance of this deterministic protocol is very similar to that of the randomized DD protocols, thus significantly outperforming the original two deterministic protocols.

---



\end{document}